\documentclass[10pt]{iopart}
\pdfoutput=1

\expandafter\let\csname equation*\endcsname=\relax
\expandafter\let\csname endequation*\endcsname=\relax
\usepackage{color}
\usepackage{amsmath}
\usepackage{amssymb}
\usepackage{enumerate}
\usepackage{graphicx}
\usepackage{mathbbol}
\usepackage{amsfonts}
\usepackage{url}
\newcommand{\nn}{\nonumber}

\newcommand{\bea}{\begin{eqnarray}}
\newcommand{\eea}{\end{eqnarray}}
\newcommand{\beq}{\begin{equation}}
\newcommand{\eeq}{\end{equation}}

\def\XXint#1#2#3{{\setbox0=\hbox{$#1{#2#3}{\int}$}
 \vcenter{\hbox{$#2#3$}}\kern-.5\wd0}}

\usepackage[utf8]{inputenc}
\usepackage{amsmath}
\usepackage{hyperref}
\usepackage{graphicx}
\usepackage{amsfonts}
\usepackage{amsthm}
\usepackage{cases}
\usepackage{bm}
\usepackage{amssymb}

\usepackage{color}
\definecolor{Blue}{rgb}{0.00, 0.00, 1.00}
\definecolor{Red}{rgb}{1.00, 0.00, 0.00}

\hypersetup{
    colorlinks=true,       % false: boxed links; true: colored links
    linkcolor=red,          % color of internal links (change box color with linkbordercolor)
    citecolor=blue,        % color of links to bibliography
    filecolor=magenta,      % color of file links
    urlcolor=cyan           % color of external links
}

\newcommand{\be}{\begin{equation}}
\newcommand{\ee}{\end{equation}}
\newcommand{\beqn}{\begin{eqnarray}}
\newcommand{\eeqn}{\end{eqnarray}}
\DeclareMathOperator{\erfc}{erfc}

\newcommand{\moy}[1]{\ensuremath{\langle #1 \rangle}}

\makeatletter
\renewcommand\@appendixstar{\@@par
 \ifnumbysec 
 \@addtoreset{table}{section}
 \@addtoreset{figure}{section}\fi
 \setcounter{section}{0}
 \setcounter{subsection}{0}
 \setcounter{subsubsection}{0}
 \setcounter{equation}{0}
 \setcounter{figure}{0}
 \setcounter{table}{0}
 \def\thesection{\Alph{section}} % this line has been \def\thesection{Appendix \Alph{section}} before
 \def\theequation{\ifnumbysec
      \Alph{section}.\arabic{equation}\else
      \Alph{section}\arabic{equation}\fi}
 \def\thetable{\ifnumbysec
      \Alph{section}\arabic{table}\else
      A\arabic{table}\fi}
 \def\thefigure{\ifnumbysec
      \Alph{section}\arabic{figure}\else
      A\arabic{figure}\fi}}
\makeatother
\begin{document}
%\title[]{Gap statistics close to the $\alpha$-quantile of a random walk}%Title of paper
\title[]{Gap statistics close to the quantile of a random walk}%Title of paper

\author{Bertrand Lacroix-A-Chez-Toine}
\address{LPTMS, CNRS, Univ. Paris-Sud, Universit\'e Paris-Saclay, 91405 Orsay, France}

\author{Satya N. Majumdar}
\address{LPTMS, CNRS, Univ. Paris-Sud, Universit\'e Paris-Saclay, 91405 Orsay, France}

\author{Gr\'egory Schehr}
\address{LPTMS, CNRS, Univ. Paris-Sud, Universit\'e Paris-Saclay, 91405 Orsay, France}

\begin{abstract}
We consider a random walk of $n$ steps starting at $x_0=0$ with a double exponential (Laplace) jump distribution. We compute exactly the distribution $p_{k,n}(\Delta)$ of the gap $d_{k,n}$ between the $k^{\rm th}$ and $(k+1)^{\rm th}$ maxima in the limit of large $n$ and large $k$, with $\alpha=k/n$ fixed. We show that the typical fluctuations of the gaps, which are of order $O( n^{-1/2})$, are described by a universal $\alpha$-dependent 
distribution, which we compute explicitly. Interestingly, this distribution has an inverse cubic tail, which implies a non-trivial $n$-dependence of the moments of the gaps. We also argue, based on numerical simulations, that this distribution is universal, i.e. it holds for more general jump distributions (not only the Laplace distribution), which are continuous, symmetric with a well defined second moment. Finally, we also compute the large deviation form of the gap distribution $p_{\alpha n,n}(\Delta)$ for $\Delta=O(1)$, which turns out to be non-universal. 
\end{abstract}

\maketitle

%\tableofcontents

\newpage

\section{Introduction}

During the last decades, extreme value statistics (EVS) have found a lot of applications in statistical physics, ranging from disordered systems \cite{Der81,BM97,DM01,PLDM03}, directed polymers and stochastic growth processes in the Kardar-Parisi-Zhang universality class \cite{baik,johann,growth,SS10,CLR10,DOT10,ACQ11,BLS12} to fluctuating interfaces \cite{RCPS01,GHPZ03,MC04,MC05}, random matrices \cite{TW94,MS2014}, random walks and Brownian motions \cite{CM2005,SM12,MMS13} all the way to cold atoms \cite{rmax}. The basic question concerns the distribution
of the maximum $x_{\max}$ (or equivalently the minium $x_{\min}$) among a collection of $N$ random variables $x_1, x_2, \cdots, x_N$, and in particular in the limit $N \to \infty$, i.e. in the thermodynamic limit. This problem is fully understood in the case of independent and identically
distributed (i.i.d.) random variables $x_i$'s, for which it is well known that there exist three distinct universality classes (Gumbel, Fr\'echet and Weibull) depending only on the tail of the parent distribution of the $x_i$'s \cite{Gumbel}. However, in many situations in statistical physics, it turns out that, often, one has to deal with {\it strongly} correlated variables \cite{satya_pal}. In fact, there exist at present very few exact results for the EVS in strongly correlated systems and it is thus crucial to identify physically relevant models for which the EVS can be computed exactly. A prototypical example of such models is the discrete-time random walk (RW), which constitutes a useful laboratory to test the effects of strong correlations on EVS \cite{Satya_Leuven}. Here we are interested in the statistics of the gaps between the consecutive maxima of a discrete-time RW.   

Indeed, the distribution of the global maximum $x_{\max}$ (or the minimum $x_{\min}$) is certainly interesting but it gives only a partial information on the system -- it concerns one single variable out of $n \gg 1$ -- and in some cases it is useful to consider the more general question
of order statistics which concern the joint statistics of the $k$-th maxima $M_{k,n}$ such that $x_{\max} = M_{1,n} > M_{2,n}> \cdots > M_{n+1,n} = x_{\min}$. Natural observables are then the gaps between successive maxima, $d_{k,n} = M_{k,n} - M_{k+1,n}$, which are useful, for instance, to quantify the phenomenon of ``crowding'' near the extremes \cite{SM07,PCMS13, RMS14,RMS15}. In physics, the statistics of the gaps were studied in the context of branching Brownian motions \cite{BD09,BD11}, as well as for noisy signals with power spectrum in $1/f^\alpha$ \cite{MOR11}, with applications in cosmology~\cite{TR77}. For RWs, such questions related to the $k$-th maximum belong to the general realm of ``fluctuation theory'' \cite{feller} and their statistics have been computed using probabilistic methods~\cite{We60, Po63,Yor95, Dassios_alpha_q, ERY95, Dassios_discrete, Chaumont}. However, much less is known about the gaps from fluctuation theory (see however~\cite{Pitman}).

Rather recently, two of us developed an independent method, based on backward equations (see below), which allowed us to solve exactly the gap statistics for random walks with symmetric exponential jumps \cite{SM12}. From this exact result, the distribution of the gaps near $x_{\max}$ (i.e., at the ``edge''), for long random walks, was obtained and it was shown to exhibit a very rich behaviour, which was conjectured to be, to a large extent,
universal, i.e. independent of the details of the jump distribution provided it has a well defined second moment. This conjecture at the ``edge'' was partly confirmed by a more recent work where it was shown that, for long RWs, the same distribution describes the typical fluctuations of the gaps of RWs with symmetric gamma-distributed jumps \cite{BMS17}. 

The goal of this paper is to reconsider this problem of the gaps of random walks and study their statistics in the ``bulk'', i.e. far away from $x_{\max}$~\footnote{Note that the terms ``bulk'' and ``edge'' are borrowed from the terminology used in Random Matrix Theory~\cite{mehta, forrester}.}, near a quantile of the random walk, for instance near the median which is at half-way between $x_{\min}$ and $x_{\max}$. We find that the gaps in the bulk also display a very rich behaviour which is quite different from the one found at the edge. We conjecture that this behaviour is also universal.

\section{Model and main results}

\begin{figure}[t]
\includegraphics[width=0.6\textwidth]{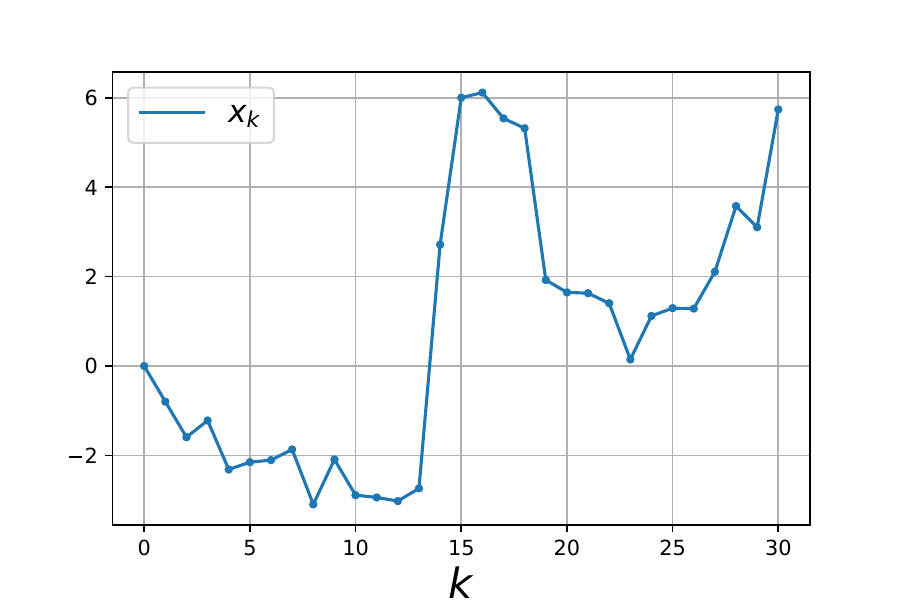}
\includegraphics[width=0.6\textwidth]{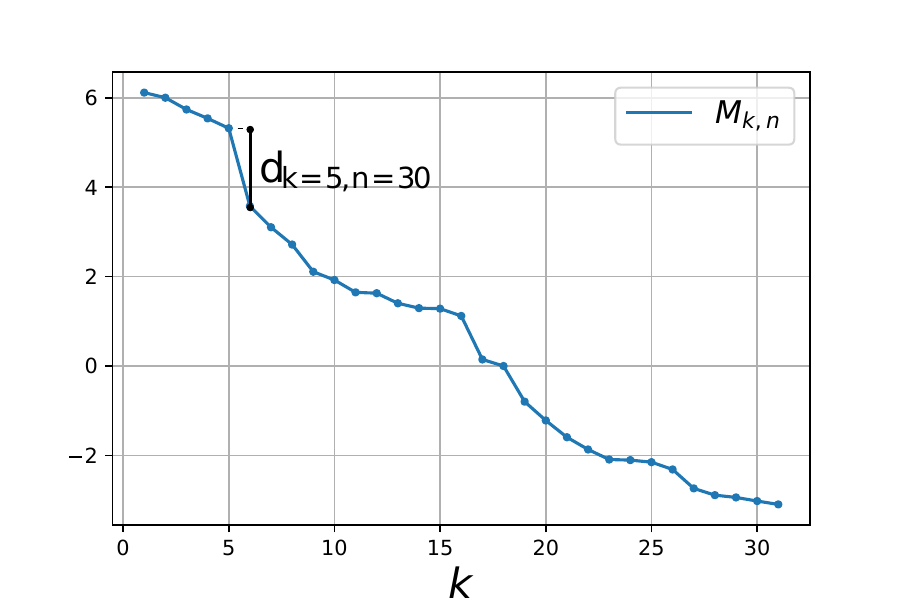}
\caption{{\bf Left}: Random walk $x_i$ of $n=30$ steps starting from $x_0=0$ with a global maximum $x_{16}=M_{1,30}$ and a global minimum $x_8=M_{31,30}$. For this realisation, one has for instance $M_{3,30}=x_{30}$ and $M_{6,30}=x_{28}$.  {\bf Right}: Values of $M_{k,n}$ as a function of $k$ for this particular realisation.}\label{Fig_walk}
\end{figure}

Let us thus consider a one-dimensional random walk in continuous space and discrete time defined by
\be\label{def_RW}
x_{i}=x_{i-1}+\eta_i\;\;{\rm for}\;\;i=1,\cdots,n\;\;\qquad {\rm starting \; from}\;\;x_0=0\;,
\ee 
where the $\eta_i$'s are i.i.d. random variables with a double exponential (or Laplace) probability distribution function (PDF) 
\be\label{exp}
f(\eta)=\frac{e^{-\sqrt{2}\frac{|\eta|}{\sigma}}}{\sqrt{2}\sigma}\;,
\ee
where $\sigma^2 = \int_{-\infty}^\infty \eta^2 \, f(\eta)\, d\eta$ is the variance of the jump distribution. Hence in the large $n$ limit the random walk in (\ref{def_RW}) converges to the Brownian motion. Note that for a walk of $n$ steps, there are $n+1$ positions $\lbrace x_0,x_1,\cdots,x_n\rbrace$. We order these positions and define the random variables $M_{k,n}$ as the $k^{\rm th}$ maximum among the positions $x_i$'s of the random walk (see Fig. \ref{Fig_walk}), such that
\be\label{Mkn_def}
M_{1,n}=x_{\max}\geq M_{2,n}\geq \cdots\geq M_{n,n}\geq M_{n+1,n}=x_{\min}\;.
\ee
Since the jump distribution is symmetric, i.e. $f(\eta)=f(-\eta)$, $x_{\max} = M_{1,n}$ has the same distribution as $-x_{\min} = -M_{n+1,n}$. Similarly $M_{2,n}$ has the same distribution as $-M_{n,n}$ and more generally $M_{k,n}$ has the same distribution as $-M_{n+2-k,n}$. Therefore  
the distribution $P_{k,n}(x)$ of $M_{k,n}$ satisfies the relation
\be
P_{k,n}(x)=P_{n+2-k,n}(-x) \;. \label{symmetry}
\ee

\begin{figure}[t]
\centering
\includegraphics[width=0.7\textwidth]{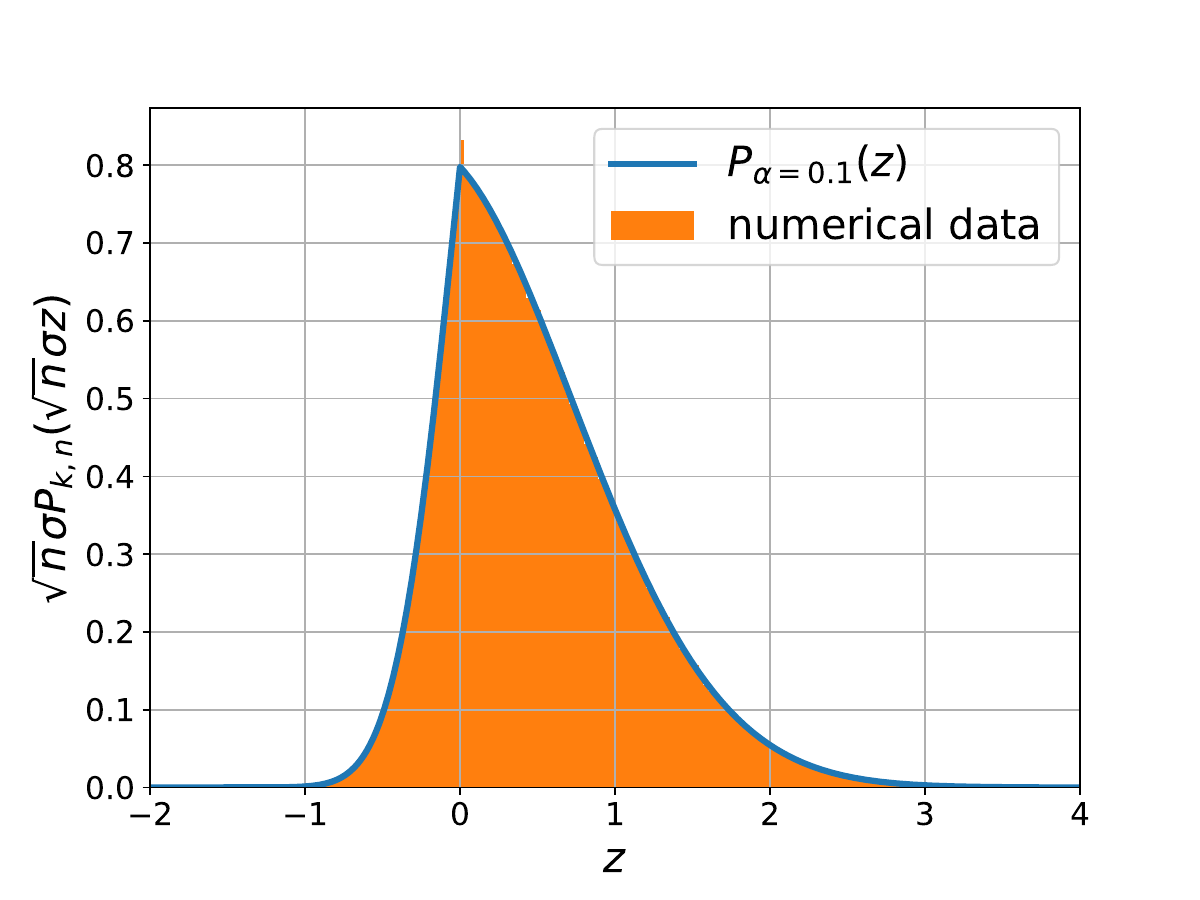}
\caption{Comparison between the rescaled PDF $\sqrt{n}\sigma P_{k,n}(\sqrt{n}\sigma z)$ of the maximum $M_{k,n}$ for $k=10^2$ and $n=10^3$ as a function of $z$ obtained from the simulation of $10^6$ random walks with Gaussian jump PDF and the scaling function $P_{\alpha}(z)$ in Eq. \eqref{P_M}. The numerical data shows a very good agreement with the analytical results.}\label{Fig_max}
\end{figure}

\noindent{\bf Distribution of the $k^{\rm th}$ maximum}. We first compute the PDF $P_{k,n}(x)$ of $M_{k,n}$ for a double exponential jump PDF, using the method introduced in \cite{SM12}. In the limit of large $n$, with $\alpha = k/n$ fixed, one finds that $P_{k = \alpha n, n}(x)$ takes the scaling form
\begin{align}
P_{k=\alpha n,n}(x)&\approx \frac{1}{\sqrt{n}\sigma}P_{\alpha}\left(\frac{x}{\sqrt{n}\sigma}\right)\label{P_M}\\
{\rm with}\;\;P_{\alpha}(z)&=\begin{cases}
\displaystyle \sqrt{\frac{2}{\pi}}e^{-\frac{z^2}{2}}\erfc\left(z\sqrt{\frac{\alpha}{2(1-\alpha)}}\right)&\;,\;\;z\geq 0\\
&\\
\displaystyle \sqrt{\frac{2}{\pi}}e^{-\frac{z^2}{2}}\erfc\left(|z|\sqrt{\frac{1-\alpha}{2\alpha}}\right)&\;,\;\;z< 0\;.
\end{cases}\nn
\end{align}
Note that the limiting distribution satisfies the relation $P_\alpha(z)=P_{1-\alpha}(-z)$ which reflects the symmetry between maxima and minima noted in Eq. (\ref{symmetry}). In Fig. \ref{Fig_max}, we plot the PDF $P_\alpha(x)$ for $\alpha =0.1 $. We see clearly that it is quite asymmetric, and from  
the exact expression (\ref{P_M}), it is easy to check that $P_{\alpha}(z) \approx e^{-z^2/(2(1-\alpha))}$ for $z \to +\infty$ while $P_{\alpha}(z) \approx e^{-z^2/(2\alpha)}$ for $z \to -\infty$. Close to $z=0$, where $P_{\alpha}(0) = \sqrt{2/\pi}$, the distribution $P_\alpha(z)$ has a cusp. From this result (\ref{P_M}), one can easily compute the first moment $\langle M_{k,n}\rangle$ in the limit of large $n$ and $k$ with $k/n = \alpha$ fixed
\bea
\langle M_{k,n}\rangle \approx \sigma\,\sqrt{n}\, {\cal M}\left( \alpha = \frac{k}{n}\right) \;, \; {\cal M}(\alpha) = \sqrt{\frac{2}{\pi}}\left(\sqrt{1-\alpha} - \sqrt{\alpha} \right) \;. \label{average}
\eea 
Note that ${\cal M}(1-\alpha) = - {\cal M}(\alpha)$, in agreement with the property in (\ref{symmetry}). In the  limit $\alpha \to 0$, one recovers that $\langle x_{\max} \rangle = \langle M_{1,n}\rangle \approx \sigma \sqrt{n} \, {\cal M}(0) = \sigma \sqrt{2\,n/\pi}$, as expected from the Brownian motion result. 

Incidentally, in this scaling limit where both $n$ and $k$ are large with $k/n = \alpha$ fixed, $M_{k,n}$ corresponds exactly to what is called, in the probability literature, the $(1-\alpha)$-quantile for  Brownian motion \cite{Yor95}. Roughly speaking, $M_{k = \alpha n,n}$ is such that a fraction $(1-\alpha)$ of the points of the trajectory of the random walks are below $M_{k=\alpha n,n}$, while a fraction $\alpha$ of the points are above it. Since in the large $n$ limit the RW converges to Brownian motion, $M_{k = \alpha n,n}/(\sigma \sqrt{n})$, converges, in the scaling limit where both $n$ and $k$ are large with $k/n = \alpha$ fixed, to the $(1-\alpha)$-quantile $q_{1-\alpha}$ of the Brownian motion,~i.e. 
\bea\label{def_alpha_q}
\frac{M_{k=\alpha n,n}}{\sigma \sqrt{n}} \underset{n\to \infty}{\longrightarrow}  q_{1-\alpha}=\inf\lbrace x:\int_0^1 \Theta(x(\tau)-x)d\tau \geq 1-\alpha\rbrace\ \;,
\eea
where $x(\tau)$ is a standard Brownian motion (with diffusion constant $D=1/2$) starting from $0$ on the time interval $[0,1]$ and ${\Theta(x)}$ is the Heaviside theta function. In fact, one can check that the formula for $P_\alpha(z)$ in Eq. (\ref{P_M}), obtained here using a backward-equation formalism, coincides with the result obtained previously in the mathematics literature using quite different probabilistic methods for Brownian motion~\cite{Yor95, Dassios_alpha_q, ERY95}. 

\vspace*{0.3cm}

\begin{figure}[t]
\centering
\includegraphics[width = 0.9\linewidth]{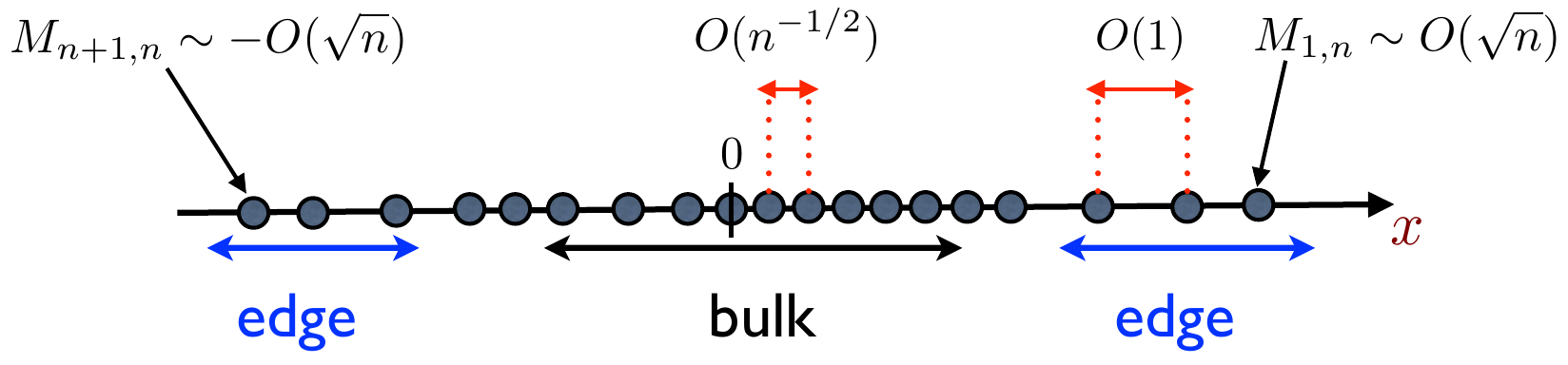}
\caption{Sketch of the point process constituted by the $k$-th maxima $M_{k,n}$ of the random walk (\ref{def_RW}) starting at $x_0=0$ after a large number of steps $n \gg 1$. At {\it the edges}, i.e. near the maximum $M_{1,n}$ and the minimum $M_{n+1,n}$ the gaps are of order $O(1)$, for large $n$, while {\it in the bulk}, i.e. far from the minimum and the maximum, the gaps are much smaller and of order $O(n^{-1/2})$ [see Eq. (\ref{mean_gap})].}\label{Fig_pp}
\end{figure}

\noindent{\bf Distribution of the $k^{\rm th}$ gap}. Our main new results concern the gaps
\be\label{def_gap}
d_{k,n}=M_{k,n}-M_{k+1,n}\geq 0\;,\;\;k=1,\cdots,n\;,
\ee
between two consecutive maxima. For a double exponential jump distribution~(\ref{exp}), the mean value of the gap $\moy{d_{k,n}}=\moy{M_{k,n}}-\moy{M_{k+1,n}}$ can be computed for any finite $n$ and $k$, yielding \cite{SM12}
\be\label{exact_gap}
\frac{\moy{d_{k,n}}}{\sigma}=\frac{\Gamma\left(k+\frac{1}{2}\right)}{\sqrt{2\pi}k!}+\frac{\Gamma\left(n-k+\frac{3}{2}\right)}{\sqrt{2\pi}(n-k+1)!} \;.
\ee
It is straightforward to extract the large $n$ behaviour of this exact expression (\ref{exact_gap}) in the two scaling regimes corresponding to $k={O}(1)$ and $k = {O}(n)$ as
\be\label{mean_gap}
\frac{\moy{d_{k,n}}}{\sigma} \approx \begin{cases}
\displaystyle \frac{\Gamma\left(k+\frac{1}{2}\right)}{\sqrt{2\pi}k!}&,\;\;n\to\infty\;,\;\;k=O(1)\\
&\\
\displaystyle\frac{\mu(\alpha)}{\sqrt{n}}&,\;\;n\to\infty\;,\;\;\alpha=\dfrac{k}{n}=O(1)\;,
\end{cases}
\ee
where the scaling function $\mu(\alpha)$ reads
\be\label{m_1}
\mu(\alpha)=\frac{1}{\sqrt{2\pi}}\left(\frac{1}{\sqrt{\alpha}}+\frac{1}{\sqrt{1-\alpha}}\right)\;.
\ee
Note that one can check that $\mu(\alpha) = -{\cal M}'(\alpha)$, where ${\cal M}(\alpha)$ is given in Eq. (\ref{average}), as expected from Eqs. (\ref{def_gap}) and (\ref{average}). This result (\ref{mean_gap}) clearly shows that, for large $n$, there are two different scales for the gaps $d_{k,n}$ depending on $k = O(1)$ or $k = O(n)$. It is useful to think about the values of the $k$-th maxima of the random walks after step $n$ as a point process on the line, as illustrated in Fig. \ref{Fig_pp}. 
Near the edges, i.e. near the maximum $x_{\max} = M_{1,n}$ and the minimum $x_{\min} = M_{n+1,n}$, the gaps are of order $O(1)$ (see the first line of Eq. (\ref{mean_gap})) while they are of order $O(n^{-1/2})$ (see the second line of Eq. (\ref{mean_gap})) in the {\it bulk}, i.e. far from $x_{\max}$ and $x_{\min}$. Note that by taking the large $k$ limit in the first line of Eq. \eqref{mean_gap} one obtains $\moy{d_{k,n}}\approx \sigma/\sqrt{2\pi k}$, while by taking the small $\alpha=k/n$ limit in the second line of Eq.~\eqref{mean_gap} one also obtains $\moy{d_{k,n}}\approx \sigma/\sqrt{2\pi \alpha n}=\sigma/\sqrt{2\pi k}$: this shows that there is a smooth matching between the edge and the bulk at the level of the first moment $\langle d_{k,n} \rangle$.% and we anticipate that this will also be the case for the full PDF of $d_{k,n}$ (and not only for the first moment). 

What about the full PDF $p_{k,n}(\Delta)$ of the gaps $d_{k,n}$ in the large $n$ limit? Near the edge, this PDF was computed for jumps with a double exponential jump distribution in Ref. \cite{SM12} and subsequently for symmetric gamma-distributed jumps in Ref.~\cite{BMS17}. In particular, it was shown in Ref. \cite{SM12} that for a double exponential jump distribution, the PDF $p_{k,n}(\Delta)$ becomes independent of $n$ in the large $n$ limit, i.e. $\lim_{n \to \infty} p_{k,n}(\Delta) = p_{k,\infty}(\Delta)$, consistent with the first line of Eq. (\ref{mean_gap}). In the large $k$ limit, it turns out \cite{SM12} that the limiting distribution $p_{k,\infty}(\Delta)$ has two different scaling behaviours, depending of $\Delta$: (i) a regime of typical fluctuations for $\Delta = O(1/\sqrt{k})$ and (ii) a large deviation regime for $\Delta = O(1)$. The most interesting result obtained in \cite{SM12} concerns the typical fluctuations 
where $p_{k,n}(\Delta)$ takes the scaling form \cite{SM12}
\be\label{scaling_k}
p_{k,n}(\Delta)\approx \frac{\sqrt{k}}{\sigma}P\left(\frac{\sqrt{k}\Delta}{\sigma}\right)
\ee
where the scaling function $P(\delta)$ is given by \cite{SM12}
\begin{eqnarray}
P(\delta)=4\left[\sqrt{\frac{2}{\pi}}(1+2\delta^2)-\delta(4\delta^2+3)e^{2\delta^2}\erfc(\sqrt{2}\delta)\right]\;. \label{typ_k_order_1}
\end{eqnarray}
%
%\be\label{scaling_k}
%p_{k,n}(\Delta)\approx \begin{cases}
%\displaystyle\frac{\sqrt{k}}{\sigma}P\left(\frac{\sqrt{k}\Delta}{\sigma}\right)&\;,\;\;\Delta=O(k^{-1/2})\\
%&\\
%\displaystyle\frac{1}{\sigma k^{3/2}}\varphi_0\left(\frac{\Delta}{\sigma}\right)&\;,\;\;\Delta=O(1)\;,
%\end{cases}
%\;,\;n\gg k\gg 1\;,
%\ee
%with the scaling functions
%\begin{align}
%&P(\delta)=4\left[\sqrt{\frac{2}{\pi}}(1+2\delta^2)-\delta(4\delta^2+3)e^{2\delta^2}\erfc(\sqrt{2}\delta)\right]\;,\label{typ_k_order_1}\\
%&\varphi_0(\tilde\Delta)=\sqrt{\frac{2}{\pi}}\frac{2+\cosh(2\sqrt{2}\tilde \Delta)}{\sinh^4(\sqrt{2}\tilde \Delta)}\;.\label{LD_k_order_1}
%\end{align}
Based on numerical simulations, it was conjectured in \cite{SM12} that the typical distribution $P(\delta)$ is universal, i.e. it does not depend on the jump distribution $f(\eta)$ as long as it is symmetric and has a finite variance $\sigma^2<\infty$. The validity of this conjecture was then reinforced by an exact analytical computation for gamma distributed jump distribution $f(\eta)=\frac{|\eta|^p}{2p!} e^{-|\eta|}$ with $p\in \mathbb{N}$ \cite{BMS17}. From this expression (\ref{typ_k_order_1}), it is easy to obtain the asymptotic behaviours of $P(\delta)$ for small and large $\delta$
\bea\label{asympt_P_edge}
P(\delta) \approx
\begin{cases}
&4 \sqrt{\dfrac{2}{\pi}} \;, \;\;\;\;\, \delta \to 0 \;, \\
&\\
&  \dfrac{3}{\sqrt{8\pi}}\dfrac{1}{\delta^4} \;, \; \delta \to \infty \;.
\end{cases}
\eea
In particular, it exhibits an interesting power law tail $P(\delta) \propto \delta^{-4}$ for large $\delta$. The large deviation regime of $p_{k,n}(\Delta)$, for $\Delta = O(1)$, can also be computed explicitly for the double exponential jump distribution (\ref{exp}) but, unlike $P(\delta)$ in (\ref{asympt_P_edge}), it turns out to be non-universal, i.e. it depends explicitly on the jump distribution \cite{SM12,BMS17} and, for this reason, it is somewhat less interesting than the typical fluctuation regime. 

%
%Surprisingly, this PDF $P(\delta)$ in Eq. \eqref{typ_k_order_1} describing the typical fluctuations of $\delta=\sqrt{k}\Delta=O(1)$ has a heavy tail,
%\be\label{tail_P}
%P(\delta)\approx \frac{3}{\sqrt{8\pi}}\frac{1}{\delta^4}\;,\;\;\delta\to\infty\;,
%\ee
%while it reaches a constant as $\delta \to 0$, $P(\delta \to 0) = 4 \sqrt{2/\pi}$. Unlike $P(\delta)$, the function $\varphi_0(\tilde \Delta)$ in (\ref{LD_k_order_1}), which describes the large deviations of the gap, is non-universal and depends on the details of the jump-distribution \cite{SM12,BMS17}.

\begin{figure}[t]
\centering
\includegraphics[width=0.6\textwidth]{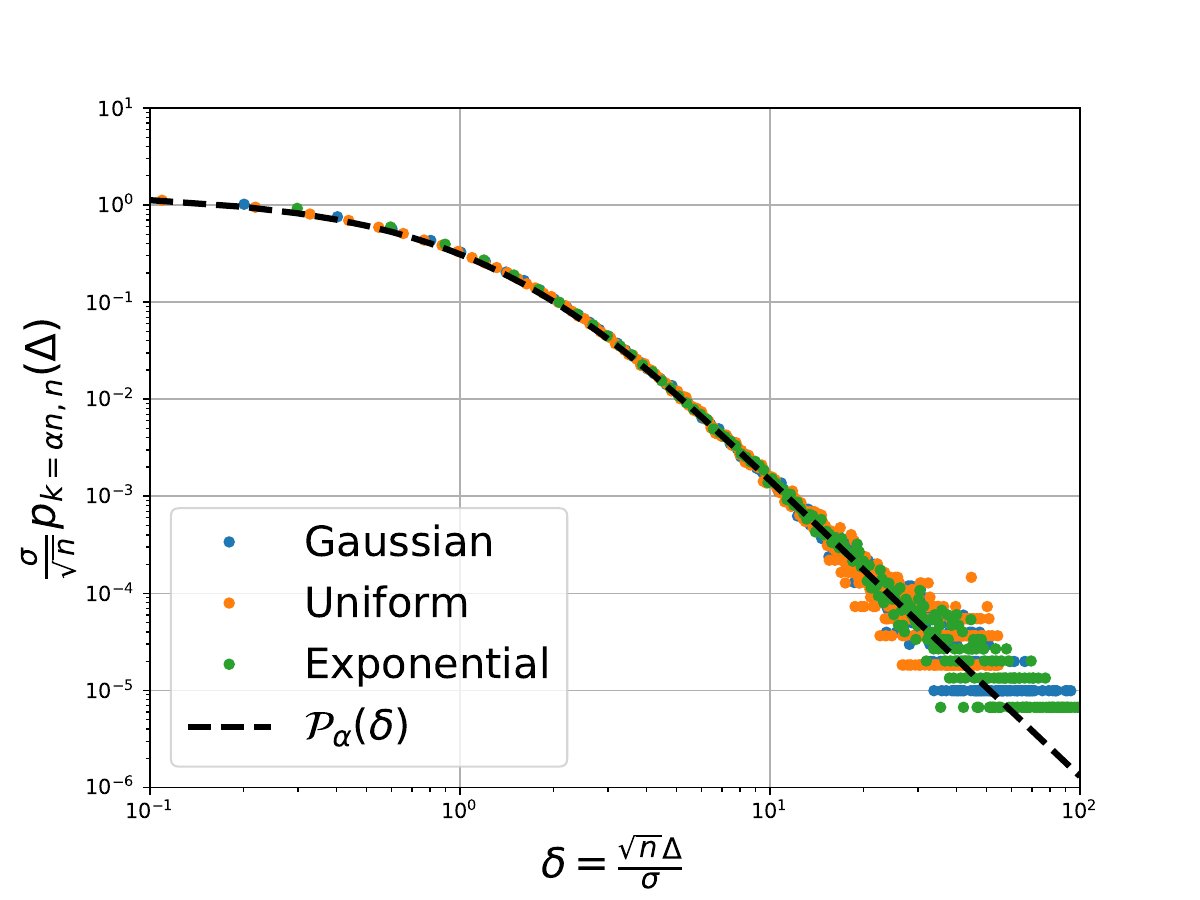}
\caption{Comparison between the rescaled PDF $\frac{\sigma}{\sqrt{n}}p_{k=\alpha n,n}(\Delta)$ of the gap $d_{k,n}$ obtained numerically for $10^6$ random walks of $n=10^3$ steps and $k=500$, hence $\alpha = k/n = 1/2$ with Gaussian (in blue), uniform (in orange) and exponential (in green) PDF of jumps $f(\eta)$ and the scaling function ${\cal P}_{\alpha=1/2}(\delta=\sqrt{n}\Delta/\sigma)$ (dashed line) given in Eq. \eqref{P_gap}, see also Eq. (\ref{P1/2}). The curves for different jumps PDF all collapse on the same master curve described by ${\cal P}_{1/2}(\delta)$, suggesting the universality of this result (\ref{P_gap}).}\label{Fig_CDF}
\end{figure}

In this paper, we derive the full PDF $p_{k,n}(\Delta)$ of the gap $d_{k,n}$ in the bulk, i.e. for large $n$ and large $k$ but keeping the ratio 
$\alpha = k/n$ fixed. We show that the behaviour in the bulk is rather different from the one found at the edge in \cite{SM12,BMS17} recalled above in Eqs.  (\ref{typ_k_order_1}) and (\ref{asympt_P_edge}), which corresponds instead to the limit $\alpha \to 0$ (i.e. $1 \ll k \ll n$). We find that this PDF $p_{k,n}(\Delta)$ again exhibits two different scaling regimes depending on $\Delta$: a typical regime for $\Delta = O(n^{-1/2})$, consistent with the second line of Eq. (\ref{mean_gap}), and a large deviation regime for $\Delta = O(1)$. Our most interesting results concern the typical regime, for $\Delta = O(n^{-1/2})$, where $p_{k,n}(\Delta)$ takes the scaling form
\be \label{P_gap_typical}
p_{k=\alpha n,n}(\Delta)\approx \displaystyle\frac{\sqrt{n}}{\sigma}{\cal P}_{\alpha}\left(\frac{\sqrt{n}\Delta}{\sigma}\right) \;,
\ee 
where the scaling function ${\cal P}_{\alpha}(\delta)$ depends continuously on the parameter $\alpha$ and is given explicitly by 
\begin{align}
{\cal P}_\alpha(\delta)=\int_{0}^{\infty} y^{2}e^{-\delta y}&\left[\frac{e^{-\frac{y^{2}}{8\alpha(1-\alpha)}}}{\pi\sqrt{\alpha(1-\alpha)}}+\frac{y\,e^{-\frac{y^{2}}{8(1-\alpha)}}}{4\sqrt{2\pi}(1-\alpha)^{\frac{3}{2}}}\erfc\left(\frac{y}{2\sqrt{2\alpha}}\right)\right.\nn\\
&\left.+\frac{y \,e^{-\frac{y^{2}}{8\alpha}}}{4\sqrt{2\pi}\alpha^{\frac{3}{2}}}\erfc\left(\frac{y}{2\sqrt{2(1-\alpha)}}\right)\right]dy\;.\label{P_gap}
\end{align}
For generic $\alpha$ this integral over $y$ can be evaluated in terms of hypergeometric functions of two variables (namely Humbert series, see Eq. (\ref{explicit_Pa_g}) in Appendix \ref{app:explicit}). For the special case $\alpha = 1/2$, which describes the gaps near the median, ${\cal P}_{1/2}(\delta)$ can be expressed in terms of elementary functions [see Eq. (\ref{P1/2})]. One can also show that in the limit $\alpha \to 0$, our result in Eqs. (\ref{P_gap_typical}) and (\ref{P_gap}) yields back the edge result  in Eqs. (\ref{typ_k_order_1}) and (\ref{asympt_P_edge}) -- this limit is however a bit subtle and is studied in detail below. Under this form (\ref{P_gap}), we notice the relation ${\cal P}_\alpha(\delta)={\cal P}_{1-\alpha}(\delta)$, which is a direct consequence of the symmetry observed above for the distribution of the $k$-th maximum in (\ref{symmetry}). In Fig. \ref{Fig_CDF}, we show a plot of this scaling function ${\cal P}_{\alpha}(\delta)$, for $\alpha = 1/2$, and compare it to numerical data (appropriately scaled according to Eq.~(\ref{P_gap_typical})) obtained by simulating random walks (\ref{def_RW}) for three different jump distributions: a double exponential PDF (\ref{exp}) -- for which our computation is exact -- but also a Gaussian PDF as well as a uniform PDF, which we can not study analytically. We find that the agreement between our theoretical results~(\ref{P_gap}) and the numerical simulations are equally good for the three jump distributions. Based on this observation, we conjecture that this distribution ${\cal P}_\alpha(\delta)$ is universal, i.e. it is independent of the jump distribution $f(\eta)$ provided it is continuous, symmetric and with a well defined variance $\sigma^2$.

From this integral representation (\ref{P_gap}) one can rather easily extract the asymptotic behaviours of the scaling function ${\cal P}_\alpha(\delta)$
\be
{\cal P}_\alpha(\delta)\approx\begin{cases}  \displaystyle4\sqrt{\frac{2}{\pi}}(\sqrt{\alpha}+\sqrt{1-\alpha}-1)&\;,\;\;\delta\to 0\;,\\
&\\
\displaystyle \frac{2}{\pi\sqrt{\alpha(1-\alpha)}}\frac{1}{\delta^3}&\;,\;\;\delta\to\infty\;.
\end{cases}\label{tail}
\ee
Interestingly, we see that the tail ${\cal P}_{\alpha}(\delta) \propto \delta^{-3}$, for finite $\alpha$ and in the bulk, is different from the tail~$\propto \delta^{-4}$ obtained at the edge [see Eq. (\ref{asympt_P_edge})]. From this inverse cubic tail one would naively conclude that the moments of the gaps (beyond the first one given in Eqs. (\ref{mean_gap}) and (\ref{m_1})) is not defined. However, this power law behaviour of the gap distribution $p_{k,n}(\Delta)$
is cut-off for $\Delta \gg n^{-1/2}$ and the higher moments are actually dominated by the large deviation regime of the PDF $p_{k,n}(\Delta)$ for $\Delta=O(1) \gg n^{-1/2}$ which we can also compute exactly (see below). The latter turns out to be non-universal. Consequently, the moments of the gaps beyond the first one, that we also study below, are to a large extent non-universal.

The paper is organised as follows. Section \ref{Mkn_sec} is dedicated to the distribution of the $k^{\rm th}$ maximum $M_{k,n}$. In Section \ref{dkn_sec} we derive the results for the gap $d_{k,n}$, which constitute our main results, before we conclude in Section \ref{sec:conclusion}. Some details of the computations have been relegated in Appendices A, B and C.

\section{Distribution of the $k^{\rm th}$ maximum}\label{Mkn_sec}

\begin{figure}[h]
\includegraphics[width=0.5\textwidth]{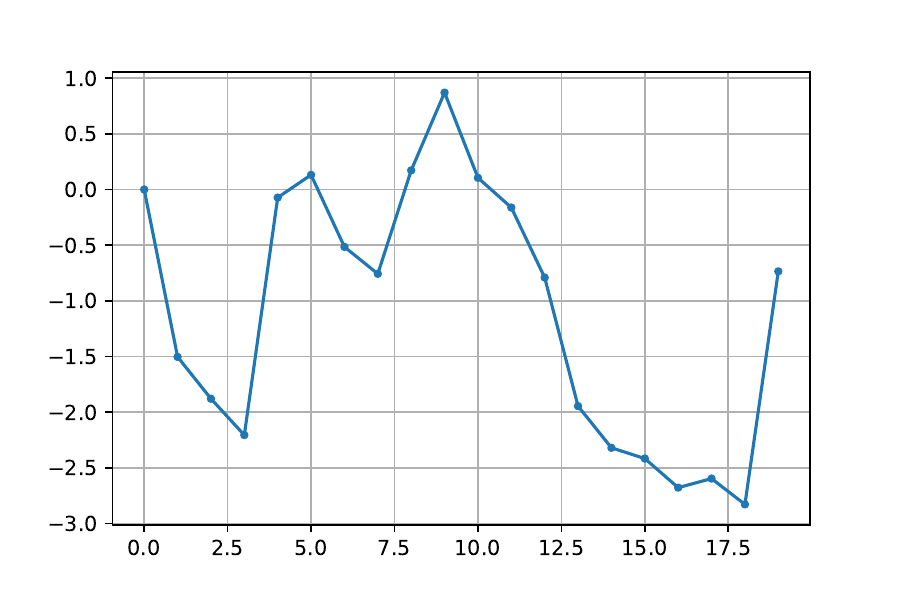}
\includegraphics[width=0.5\textwidth]{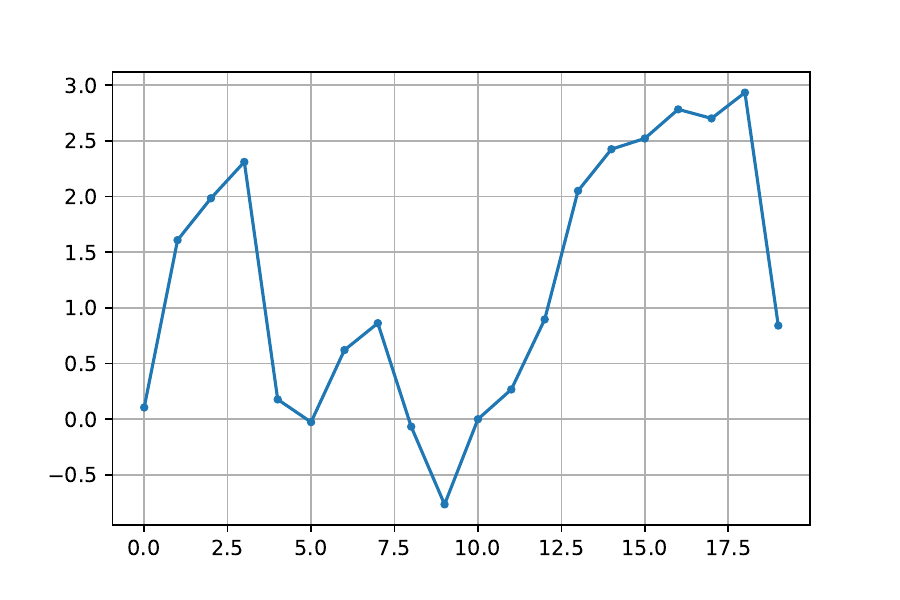}
\caption{{\bf Left}: Random walk $x_i$ of $n=20$ steps starting from $x_0=0$ with jump distribution $f(\eta)$. For this walk, the $4^{\rm th}$ maximum $M_{4,20}=x_{10}\geq 0$. {\bf Right}: Affine transformation of the random walk $y_i=x_{10}-x_i$ starting from $y_0=x_{10}$. This walk has the same jump distribution $f(\eta)$ and has exactly $4$ points below $y=0$.}\label{Fig_path}
\end{figure}

We first expose a method which allows us to obtain the distribution of $M_{k,n}$ that we will generalise in the next section to obtain the distribution of the gaps $d_{k,n}$. It relies on the identity for the cumulative distribution function (CDF) of the $k^{\rm th}$ maximum $M_{k,n}$
\be
F_{k,n}(x)={\rm Prob.}\left[M_{k,n}\leq x\right]={\rm Prob.}\left[N_x\leq k\right]\;,
\ee 
where $N_x$ is the counting process for the number of steps where the random walk takes values above $x$. Indeed, there are exactly $k$ positions among the $n+1$ positions $x_i$'s of the walk such that $x_{i}\leq M_{k,n}$ (see Fig. \ref{Fig_walk} for an example). Note that this identity remains valid for any discrete time stochastic process.
We introduce the probability $q_{k,n}(x)$ that a random walk of $n$ steps starting at $x_0=x$ has exactly $k$ points on the negative axis between step $1$ and step $n$, i.e. $N_0=n-k$ for the walk starting from $x_0=x$. The CDF $F_{k,n}(x)$ of $M_{k,n}$ can then be expressed from an elementary path transformation (see Fig. \ref{Fig_path}) as \cite{SM12}
\be\label{Q_k_n}
F_{k,n}(x)=\begin{cases}
\displaystyle\sum_{l=0}^{k-1} q_{l,n}(x)&\;,\;\;x\geq 0\\
&\\
\displaystyle\sum_{l=0}^{k-2} q_{n-l,n}(-x)&\;,\;\;x<0\;.
\end{cases}
\ee

The probability $q_{k,n}(x)$ can be constructed recursively, using the equation \cite{SM12}
\be\label{q_k_n_1}
q_{k,n}(x)=\int_0^{\infty}dx' f(x'-x)q_{k,n-1}(x')+\int_{-\infty}^0 dx' f(x'-x)q_{n-k,n-1}(-x')\;,
\ee
together with the initial condition $q_{0,0}(x)=1$ and $q_{k,n}(x)=0$ for $k>n$.
The first term of this equation describes the case where the walk has an additional initial jump from $x>0$ to $x'>0$, while the second term describes a jump from $x>0$ to $x'<0$. To solve this equation (\ref{q_k_n_1}) it is useful \cite{SM12} to introduce the auxiliary function $r_{k,n}(x) = q_{n-k,n}(x)$ which is the probability that a RW starting from $x_0=x$ has $k$ points above $0$ between step $1$ and step $n$. One can then write two coupled equations for $q_{k,n}(x)$ and $r_{k,n}(x)$~\cite{SM12}
\begin{align}\label{coupled_rq}
&q_{k,n}(x) = \int_0^{\infty}dx' f(x'-x)q_{k,n-1}(x')+\int_0^{\infty} dx' f(x+x')r_{k-1,n-1}(x') \\
&r_{k,n}(x) = \int_0^{\infty}dx' f(x'-x)r_{k-1,n-1}(x') + \int_{0}^\infty dx' f(x+x') q_{k,n-1}(x')
\end{align}
These integral equations (\ref{coupled_rq}) can be solved by generating function techniques. We introduce 
\be
\tilde q(z,s;x)=\sum_{n= 0}^{\infty}\sum_{k=0}^{n} s^n z^k q_{k,n}(x)\;\;{\rm and}\;\;\tilde r(z,s;x)=\sum_{n= 0}^{\infty}\sum_{k=0}^{n} s^n z^k q_{n-k,n}(x)\;,
\ee
and obtain from (\ref{coupled_rq}) the set of coupled integral equations
\begin{align}
\tilde q(z,s;x)&=1+s\int_0^{\infty}dx' f(x'-x)\tilde q(z,s;x')+zs\int_{0}^{\infty} dx' f(x'+x)\tilde r(z,s;x')\label{q_tilde}\\
\tilde r(z,s;x)&=1+zs\int_0^{\infty}dx' f(x'-x)\tilde r(z,s;x')+s\int_{0}^{\infty} dx' f(x'+x)\tilde q(z,s;x')\;.\label{r_tilde}
\end{align}
These equations, valid for any distribution of jumps $f(\eta)$, turn out to be very difficult to solve in general. However, for a double exponential jump distribution (\ref{exp})
%\be\label{exp}
%f(\eta)=\frac{e^{-\sqrt{2}\frac{|\eta|}{\sigma}}}{\sqrt{2}\sigma}\;,
%\ee
they can be solved exactly using the identity $f''(\eta)=\frac{2}{\sigma^2} \left[f(\eta)-\delta(\eta)\right]$. Differentiating twice Eqs.~\eqref{q_tilde} and \eqref{r_tilde} with respect to $x$, we obtain two decoupled differential equations
\begin{align}
\frac{\sigma^2}{2}\partial_{x}^2\tilde q(z,s;x)&=(1-s)\tilde q(z,s;x)-1 \label{q_tilde_2}\\
\frac{\sigma^2}{2}\partial_{x}^2\tilde r(z,s;x)&=(1-zs)\tilde r(z,s;x)-1\;.\label{r_tilde_2}
\end{align}
Discarding the diverging solution for $x\to +\infty$, we obtain
\be
\tilde q(z,s;x)=a(z,s)e^{-\sqrt{2(1-s)}\frac{x}{\sigma}}+\frac{1}{1-s}\;,\;\;\tilde r(z,s;x)=b(z,s)e^{-\sqrt{2(1-zs)}\frac{x}{\sigma}}+\frac{1}{1-zs}\;.
\ee
The values of $a(z,s)$ and $b(z,s)$ are obtained by substituting back these forms in Eqs. \eqref{q_tilde_2} and \eqref{r_tilde_2}. 
Finally, the solutions read
\begin{subequations}\label{GF1}
\begin{align}
&\tilde q(z,s;x)=\left(\frac{1}{\sqrt{1-zs}}-\frac{1}{\sqrt{1-s}}\right)\frac{e^{-\sqrt{2(1-s)}\frac{x}{\sigma}}}{\sqrt{1-s}}+\frac{1}{1-s}\;,\\
&\tilde r(z,s;x)=\left(\frac{1}{\sqrt{1-s}}-\frac{1}{\sqrt{1-zs}}\right)\frac{e^{-\sqrt{2(1-zs)}\frac{x}{\sigma}}}{\sqrt{1-zs}}+\frac{1}{1-zs}\;.
\end{align}
\end{subequations}
The generating function $\tilde P(z,s;x)$ of the PDF $P_{k,n}(x)=F_{k,n}'(x)$ can be worked out explicitly in terms of $\tilde q(z,s;x)$ and $\tilde r(z,s;x)$ using Eq. \eqref{Q_k_n}
\be\label{GF2}
\tilde P(z,s;x)=\sum_{n=0}^{\infty}\sum_{l=0}^{n}s^n z^k P_{k,n}(x)=\begin{cases}
\displaystyle\frac{z}{1-z}\partial_x \tilde q(z,s;x)&\;,\;\;x\geq 0\\
&\\
\displaystyle\frac{z^2}{1-z}\partial_x \tilde r(z,s;-x)&\;,\;\;x<0\;.
\end{cases}
\ee
To obtain the large $n$ behaviour of $P_{k,n}(x)$, we perform a change of variables in the generating function $\tilde P(z,s;x)$ and set $s=\exp(-p)$ and $z=\exp(-q)$ with $p\sim q\ll 1$. In this limit, the discrete sums over $n$ and $k$ can be replaced by integrals. The generating function $\tilde P(z,s;x)$ therefore converges towards the double Laplace transform of the PDF $P_{k,n}(x)$ with respect to $n$ and $k$,
\be
\tilde P(z=e^{-q},s=e^{-p};x)\approx \tilde \Pi(q,p;x)=\int_0^{\infty}dn\int_0^{\infty}dk \, e^{-p n-kq} P_{k,n}(x)\;,
\ee 
where the function $\tilde \Pi(q,p;x)$ can be computed from Eqs. (\ref{GF1}) and (\ref{GF2}). It reads, at leading order for $p\sim q\ll 1$,
\be
\tilde \Pi(q,p;x)\approx \begin{cases}
\displaystyle\frac{\sqrt{2}}{\sigma}\left(\frac{1}{\sqrt{p}}-\frac{1}{\sqrt{p+q}}\right)\frac{e^{-\sqrt{2p}\frac{x}{\sigma}}}{q}&\;,\;\;x\geq 0\\
&\\
\displaystyle\frac{\sqrt{2}}{\sigma}\left(\frac{1}{\sqrt{p}}-\frac{1}{\sqrt{p+q}}\right)\frac{e^{\sqrt{2(p+q)}\frac{x}{\sigma}}}{q}&\;,\;\;x<0\;.
\end{cases}
\ee
Using the inverse Laplace transforms
\be
{\cal L}^{-1}_{u\to \tau}\left(e^{-\sqrt{2(u+v)}x}\right)=\frac{x\;e^{-\frac{x^2}{2\tau}}e^{-v\tau}}{\sqrt{2\pi}\tau^{3/2}}\;,\;\;{\cal L}^{-1}_{u\to \tau}\left(\frac{1}{\sqrt{u+v}}\right)=\frac{e^{-v\tau}}{\sqrt{\pi \tau}}\;,
\ee
we invert the Laplace transform from $p$ to $n$, yielding
\be
{\cal L}^{-1}_{p\to n}\left(\tilde \Pi(q,p;x)\right)\approx \begin{cases}
\displaystyle\frac{1}{\sigma}\int_0^{n}d\tau \frac{x\;e^{-\frac{x^2}{2\sigma^2 \tau}}}{\pi\sigma\sqrt{n-\tau}\tau^{3/2}}\frac{1-e^{-q(n-\tau)}}{q}&\;,\;\;x\geq 0\\
&\\
\displaystyle\frac{1}{\sigma}\int_0^{n}d\tau \frac{|x|\;e^{-\frac{x^2}{2\sigma^2 \tau}}}{\pi\sigma\sqrt{n-\tau}\tau^{3/2}}\frac{e^{-q\tau}-e^{-qn}}{q}&\;,\;\;x<0\;.
\end{cases}
\ee
Finally, using the identity
\be
{\cal L}^{-1}_{v\to t}\left(\frac{e^{-v\tau}}{v}\right)=\Theta(t-\tau)\;,
\ee
we invert the Laplace transform from $q$ to $k$, yielding
\be
P_{k,n}(x)\approx \begin{cases}
\displaystyle\frac{1}{\sigma}\int_0^{n-k}d\tau \frac{x\;e^{-\frac{x^2}{2\sigma^2 \tau}}}{\pi\sigma\sqrt{n-\tau}\tau^{3/2}}&\;,\;\;x\geq 0\\
&\\
\displaystyle\frac{1}{\sigma}\int_0^{k}d\tau \frac{|x|\;e^{-\frac{x^2}{2\sigma^2 \tau}}}{\pi\sigma\sqrt{n-\tau}\tau^{3/2}}&\;,\;\;x<0\;.
\end{cases}
\ee 
Changing the variable $\tau \to n\tau$ in this expression, it takes the scaling form described in the first line of Eq. \eqref{P_M}, where the scaling function $P_{\alpha}(z)$ is given as
\be\label{P_a_1}
P_{\alpha}(z)\approx \begin{cases}
\displaystyle \int_0^{1-\alpha}d\tau \frac{z\;e^{-\frac{z^2}{2\tau}}}{\pi\sqrt{1-\tau}\tau^{3/2}}&\;,\;\;z\geq 0\\
&\\
\displaystyle \int_0^{\alpha}d\tau \frac{|z|\;e^{-\frac{z^2}{2\tau}}}{\pi\sqrt{1-\tau}\tau^{3/2}}&\;,\;\;z<0\;.
\end{cases}
\ee 
The integral in Eq. \eqref{P_a_1} can be computed exactly using the identity
\be
\partial_t\left[\erfc\left(z\sqrt{\frac{1-t}{2t}}\right)\right]=\frac{z e^{-(1-t)z^2/2t}}{\sqrt{2\pi(1-t)}t^{3/2}}\;,
\ee
where $\erfc(x)=\frac{2}{\sqrt{\pi}}\int_x^{\infty}e^{-u^2}du$ is the complementary error function, leading to the final expression of $P_{\alpha}(z)$ in the second line of Eq. \eqref{P_M}.

We conclude this section by mentioning that there is an alternative method to obtain the PDF of $M_{k,n}$ for a discrete time random walk. It was obtained in the limit of large $n$ and for $\alpha=k/n=O(1)$ in Ref. \cite{Dassios_discrete}, making use of the identity in law derived in \cite{We60} and extended in \cite{Po63} (see also the more recent work \cite{Chaumont})
\be\label{id_max_min}
M_{k,n}\equiv\max_{0\leq i\leq n+1-k} x_i+\min_{0\leq j\leq k-1} x_j'\;,
\ee
where $\lbrace x_i\rbrace$ and $\lbrace x_i'\rbrace$ are two independent random walks with same jump distribution $f(\eta)$ starting at $x_0=x_0'=0$. This alternative method can be exploited for any jump distribution $f(\eta)$, even if $\sigma^2=\infty$. The distribution of $M_{k,n}$ was shown to take universal scaling forms, depending on the behaviour of $\hat f(k)=\int_{-\infty}^{\infty}\frac{dk}{2\pi}e^{-i k\eta}f(\eta)$ for small $k$. However, to our knowledge, there is no direct extension of this method to compute the distribution of the gap $d_{k,n}$. In the next section, we will show 
how to extend the method presented in this section to obtain exact results for the PDF $p_{k,n}(\Delta)$ of the gap $d_{k,n}$ for a double exponential jump distribution.

\section{Distribution of the $k^{\rm th}$ gap}\label{dkn_sec}
Our starting point is the joint CDF of $M_{k,n}$ and $M_{k+1,n}$,
\be
S_{k,n}(x,y)={\rm Prob.}\left[M_{k,n}\geq y,M_{k+1,n}\leq x\right]\;,
\ee
from which the PDF of the gap $d_{k,n}$ can be obtained as
\be\label{PDF_gap_1}
p_{k,n}(\Delta)=-\int dx \int dy \partial_{xy}^2 S_{k,n}(x,y)\Theta(y-x)\delta(\Delta-y+x)\;.
\ee
To obtain the joint CDF $S_{k,n}(x,y)$, we introduce the probability $Q_{k,n}(x,\Delta)$ that a random walk of $n$ steps starting at $x_0=x$ has exactly $k$ points below $0$ and no point in the interval $[-\Delta,0]$ between step $1$ and step $n$. Using a simple path transformation, one obtains the relation \cite{SM12}
\be\label{S_Q}
S_{k,n}(x,y)=\begin{cases}
Q_{k,n}(x,y-x)&\;,\;\;x>0\\
&\\
0&\;,\;\;x<0\;\;{\rm and}\;\;y>0\\
&\\
Q_{n+1-k,n}(-y,y-x)&\;,\;\;x<0\;\;{\rm and}\;\;y<0\;.
\end{cases}
\ee
The probability $Q_{k,n}(x,\Delta)$ can be obtain recursively using the relation
\be
Q_{k,n}(x,\Delta)=\int_0^{\infty}dx' f(x-x')Q_{k,n-1}(x',\Delta)+\int_{-\infty}^{0}dx' f(x-x'+\Delta)Q_{n-k,n-1}(-x',\Delta)\;.
\ee
This is a similar recursion relation as for $q_{k,n}(x)$ in Eq. \eqref{q_k_n_1}. 
Therefore, introducing the generating functions
\be
\tilde Q(z,s;x,\Delta)=\sum_{n=0}^{\infty}\sum_{k=0}^{n}s^n z^k Q_{k,n}(x,\Delta)\;,\;\;\tilde R(z,s;x,\Delta)=\sum_{n=0}^{\infty}\sum_{k=0}^{n}s^n z^k Q_{n-k,n}(x,\Delta)\;,
\ee 
they satisfy a set of coupled integral equations very similar to Eqs. \eqref{q_tilde} and \eqref{r_tilde} \cite{SM12}
\begin{eqnarray}
\tilde Q(z,s;x,\Delta)&=&1+s\int_0^{\infty}dx' f(x-x')\tilde Q(z,s;x',\Delta) \nonumber \\
&+&zs\int_0^{\infty}dx' f(x+x'+\Delta)\tilde R(z,s;x',\Delta) \;, \label{inteq_1}\\
\tilde R(z,s;x,\Delta)&=&1+zs\int_0^{\infty}dx' f(x-x')\tilde R(z,s;x',\Delta) \nonumber \\
&+&s\int_0^{\infty}dx' f(x+x'+\Delta)\tilde Q(z,s;x',\Delta)\label{inteq_2}\;.
\end{eqnarray}
For a double exponential jump distribution $f(\eta)$ as in Eq. \eqref{exp}, these integral equations can be recast as differential equations \cite{SM12, BMS17}: indeed one can easily show that the generating functions follow the set of differential equations \eqref{q_tilde_2} and \eqref{r_tilde_2}, with the substitutions $\tilde q \to \tilde Q$ and $\tilde r \to \tilde R$. Solving these equations, we obtain
\begin{align}
&\tilde Q(z,s;x,\Delta)=A_1(z,s;\Delta)e^{-\sqrt{2(1-s)}\frac{x}{\sigma}}+\frac{1}{1-s}\;,\\
&\tilde R(z,s;x,\Delta)=B_1(z,s;\Delta)e^{-\sqrt{2(1-zs)}\frac{x}{\sigma}}+\frac{1}{1-zs}\;.
\end{align}
The coefficients $A_1$ and $B_1$ are then determined by inserting these solutions back in Eqs. \eqref{inteq_1} and \eqref{inteq_2}. This yields
\be
A_1(z,s;\Delta)=\frac{\frac{zs}{\sqrt{1-zs}}-\frac{s}{1-s}\left[\sqrt{1-zs}\cosh\left(\frac{\sqrt{2}\Delta}{\sigma}\right)+\sinh\left(\frac{\sqrt{2}\Delta}{\sigma}\right)\right]}{(\sqrt{(1-zs)(1-s)}+1)\sinh\left(\frac{\sqrt{2}\Delta}{\sigma}\right)+(\sqrt{1-zs}+\sqrt{1-s})\cosh\left(\frac{\sqrt{2}\Delta}{\sigma}\right)}\;,
\ee
and $B_1(z,s;\Delta)=A_1(z^{-1},zs;\Delta)$. From Eqs.  \eqref{PDF_gap_1} and \eqref{S_Q}, we can express the generating function $\tilde p(z,s;\Delta)=\sum_{n=0}^{\infty}\sum_{k=0}^{n}s^n z^k p_{k,n}(\Delta)$, in terms of the coefficients $A_1$ and $B_1$ (see Appendix A of Ref. \cite{BMS17} for more details). This yields 
\begin{align}
\tilde p(z,s;\Delta)=&\partial_{\Delta}A_1(z,s;\Delta)+\frac{\sigma}{\sqrt{2(1-s)}}\partial_{\Delta}^2 A_1(z,s;\Delta)\label{P_A_B}\\
&+z e^{\sqrt{2(1-zs)}\frac{\Delta}{\sigma}}\left(\partial_{\Delta}B_1(z,s;\Delta)+\frac{\sigma}{\sqrt{2(1-zs)}}\partial_{\Delta}^2 B_1(z,s;\Delta)\right)\;.\nn
\end{align}
As in the case of the PDF of $M_{k,n}$, we are interested in the limit $n\to \infty$ and $k\to \infty$, which is conveniently obtained by performing the changes of variables $z=e^{-q}$ and $s=e^{-p}$ and by taking the limit $p,q \to 0$. In this limit, the discrete sums over $k$ and $n$ can then be replaced by integrals, yielding
\be
\tilde p(z=e^{-q},s=e^{-p};\Delta)\approx \tilde \pi(p+q,p;\Delta)=\int_0^{\infty}d(n-k)\int_0^{\infty}dk \,e^{-p(n-k) -(p+q)k}p_{k,n}(\Delta)\;,
\ee
where $\tilde \pi(p+q,p;\Delta)$ is the double Laplace transform of the PDF $p_{k,n}(\Delta)$ with respect to $k$ and $n-k$. To simplify the notations, we denote from now on $r=p+q$. Note that we anticipate that $\tilde \pi(r=p+q,p;\Delta)$ is symmetric in $p$ and $p+q$, since $p_{k,n}(\Delta) = p_{n-k,n}(\Delta)$. We will now analyse the PDF in the large $n$ limit and treat separately the typical fluctuations for $\Delta=O(n^{-1/2})$ and the atypically large fluctuations for $\Delta=O(1)$.

\subsection{Typical regime of fluctuation}

 To analyse the typical regime, we need to obtain the behaviour of $A_1(z=e^{-q},s=e^{-p};\Delta)$ (resp. $B_1$) in the regime $p\sim q\sim \Delta^2$. In this regime, the coefficients take the scaling form
\begin{align}
A_1(z=e^{-q},s=e^{-p};\Delta)&\approx a_1(r=p+q,p;\Delta)=-\frac{r-p+\frac{\sqrt{2r}\Delta}{\sigma}}{p\sqrt{r}\left(\sqrt{p}+\sqrt{r}+\frac{\sqrt{2}\Delta}{\sigma}\right)}\;,\\
B_1(z=e^{-q},s=e^{-p};\Delta)&\approx b_1(r=p+q,p;\Delta)=-\frac{p-r+\frac{\sqrt{2p}\Delta}{\sigma}}{r\sqrt{p}\left(\sqrt{p}+\sqrt{r}+\frac{\sqrt{2}\Delta}{\sigma}\right)}\;.
\end{align}
Note that $b_1(r,p;\Delta)=a_1(p,r;\Delta)$. Inserting these expressions in Eq. \eqref{P_A_B}, we realise that the leading terms are the second derivatives with respect to $\Delta$, as $\Delta$ is small and $\sqrt{1-s}\sim \sqrt{p}\ll 1$. The scaling function $\tilde \pi(r=p+q,p;\Delta)$ thus reads in this limit
\begin{align}
\tilde \pi(r,p;\Delta)&\approx\frac{\sigma}{\sqrt{2}}\partial_{\Delta}^2\left(\frac{a_1(r,p;\Delta)}{\sqrt{p}}+\frac{a_1(p,r;\Delta)}{\sqrt{r}}\right)\\
&=\frac{2\sqrt{2}}{\sigma\left(\sqrt{r}+\sqrt{p}+\frac{\sqrt{2}\Delta}{\sigma}\right)^3}\left(\frac{1}{\sqrt{r}}+\frac{1}{\sqrt{p}}\right)^2\;.
\end{align}
It is symmetric in $p$ and $r=p+q$, reflecting the symmetry of the PDF $p_{k,n}(\Delta)$ in $k$ and $n-k$. To invert the Laplace transforms with respect to $r$ and $s$ we first use the identity
\be
\frac{2}{(x+p)^3}=\int_0^{\infty}y^2 e^{-y(x+p)}dy\;,
\ee
to obtain
\be
\tilde \pi(r,p;\Delta)=\frac{1}{2\sigma}\int_0^{\infty}y^2 e^{-y\left(\frac{\Delta}{\sigma}+\sqrt{r/2}+\sqrt{p/2}\right)}\left(\frac{1}{\sqrt{r}}+\frac{1}{\sqrt{p}}\right)^2 dy\;.
\ee
Finally, using the Laplace inversion formulae
\bea
&&{\cal L}^{-1}_{u\to \tau}\left(e^{-\sqrt{u}x}\right)=\frac{x e^{-\frac{x^2}{4\tau}}}{2\sqrt{\pi}\tau^{3/2}}\;,\;\;{\cal L}^{-1}_{u\to \tau}\left(\frac{e^{-\sqrt{u}x}}{\sqrt{u}}\right)=\frac{e^{-\frac{x^2}{4\tau}}}{\sqrt{\pi\tau}}\;,\\
&&{\cal L}^{-1}_{u\to \tau}\left(\frac{e^{-\sqrt{u}x}}{u}\right)=\erfc\left(\frac{x}{2\sqrt{\tau}}\right)\;,
\eea
we obtain the PDF
\begin{align}
p_{k,n}(\Delta)\approx\int_{0}^{\infty} y^{2}e^{-\frac{\Delta}{\sigma}y}&\left[\frac{e^{-\frac{n y^{2}}{8k(n-k)}}}{\pi\sqrt{k(n-k)}}+\frac{ye^{-\frac{y^{2}}{8(n-k)}}}{4\sqrt{2\pi}(n-k)^{\frac{3}{2}}}\erfc\left(\frac{y}{2\sqrt{2k}}\right)\right.\nn\\
&\left.+\frac{y e^{-\frac{y^{2}}{8k}}}{4\sqrt{2\pi}k^{\frac{3}{2}}}\erfc\left(\frac{y}{2\sqrt{2(n-k)}}\right)\right]dy\;.\label{PDF_gap_unscaled}
\end{align}
Performing the change of variable $y\to y/\sqrt{n}$ in Eq. \eqref{PDF_gap_unscaled}, we eventually obtain that $p_{k,n}(\Delta)$ 
takes the scaling form announced in Eq. \eqref{P_gap_typical} with the scaling function given in Eq. \eqref{P_gap} that we reproduce here
\begin{align}\label{eq_p_text}
{\cal P}_\alpha(\delta)=\int_{0}^{\infty} y^{2}e^{-\delta y}&\left[\frac{e^{-\frac{y^{2}}{8\alpha(1-\alpha)}}}{\pi\sqrt{\alpha(1-\alpha)}}+\frac{ye^{-\frac{y^{2}}{8(1-\alpha)}}}{4\sqrt{2\pi}(1-\alpha)^{\frac{3}{2}}}\erfc\left(\frac{y}{2\sqrt{2\alpha}}\right)\right.\nn\\
&\left.+\frac{y e^{-\frac{y^{2}}{8\alpha}}}{4\sqrt{2\pi}\alpha^{\frac{3}{2}}}\erfc\left(\frac{y}{2\sqrt{2(1-\alpha)}}\right)\right]dy\;.
\end{align}
For generic $\alpha$ this integral has a rather complicated expression in terms of hypergeometric functions of two variables (namely Humbert series, see Eq. (\ref{explicit_Pa_g}) in Appendix \ref{app:explicit}). However, for $\alpha = 1/2$ it has an explicit expression in terms of elementary functions given in (\ref{P1/2}). 

From this expression (\ref{eq_p_text}), we can compute the mean of the distribution, recovering the result of Eq. \eqref{m_1} (see Appendix \ref{m_1_P_a} for details). However, this distribution has a heavy tail as seen in Eq. \eqref{tail} and its moments of order $p\geq 2$ are infinite. In Fig. \ref{Fig_CDF}, we compare the scaling function ${\cal P}_\alpha(\delta)$ to numerical results obtained for $10^6$ simulations of random walks of $n=10^3$ steps with exponential, Gaussian and uniform distribution of jump $f(\eta)$, suggesting the universality of the result.

%\vspace*{0.1cm}
\noindent{\bf The limit $\alpha \to 0$}. We first check that in the limit $\alpha \to 0$, this distribution ${\cal P}_\alpha(\delta)$ yields back the result at the edge obtained in Ref. \cite{SM12}, given in Eqs. \eqref{scaling_k} and \eqref{typ_k_order_1}. To recover this edge result, we need to take simultaneously the limit 
$\alpha=k/n\to 0$ and $\delta=\sqrt{n}\Delta\to \infty$ but keeping $\sqrt{\alpha}\,\delta= \sqrt{k} \,\Delta$ fixed [see Eq. (\ref{typ_k_order_1})]. In this scaling limit, we show that ${\cal P}_\alpha(\delta)$ in Eq. (\ref{P_gap}) takes the scaling form
\be\label{small_alpha}
{\cal P}_\alpha(\delta)\approx\sqrt{\alpha}P(\sqrt{\alpha}\delta)\;,
\ee
where $P(\delta)$ is given in Eq. \eqref{typ_k_order_1}. To show this result (\ref{small_alpha}), we demonstrate equivalently, setting $\delta = \delta'/\sqrt{\alpha}$ with $\delta'$ fixed, that 
\be \label{demo}
\lim_{\alpha \to 0} \frac{1}{\sqrt{\alpha}}{\cal P}_{\alpha}\left(\frac{\delta'}{\sqrt{\alpha}}\right) = P(\delta') \;.
\ee
To show (\ref{demo}) we write ${\cal P}_\alpha(\delta'/\sqrt{\alpha})$ starting from Eq. (\ref{eq_p_text}) and perform the change of variable 
$y\to z=y/\sqrt{\alpha}$ to obtain
\begin{align}\label{alpha_0}
\frac{1}{\sqrt{\alpha}}{\cal P}_{\alpha}\left(\frac{\delta'}{\sqrt{\alpha}}\right)=&\int_{0}^{\infty} z^{2}e^{-\delta' z}\left[\frac{\sqrt{\alpha }e^{-\frac{z^{2}}{8(1-\alpha)}}}{\pi\sqrt{(1-\alpha)}}+\frac{z\alpha^{3/2} e^{-\frac{\alpha z^{2}}{8(1-\alpha)}}}{4\sqrt{2\pi}(1-\alpha)^{\frac{3}{2}}}\erfc\left(\frac{z}{2\sqrt{2}\alpha}\right)\right.\\
&\left.+\frac{z e^{-\frac{z^{2}}{8}}}{4\sqrt{2\pi}}\erfc\left(\frac{z\sqrt{\alpha}}{2\sqrt{2(1-\alpha)}}\right)\right]dz\approx \int_{0}^{\infty}\frac{z^3 e^{-\frac{z^{2}}{8}-\delta' z}}{4\sqrt{2\pi}}dz\;,\;\;\alpha\to 0\;, \nn
\end{align}
since only the last term (in the integrand) survives in the limit $\alpha \to 0$. Finally, evaluating explicitly the remaining integral over $z$ in Eq. \eqref{alpha_0} yields back the scaling function $P(\delta')$ given in Eq. \eqref{typ_k_order_1}. This shows the scaling form in Eq. \eqref{small_alpha}.

\begin{figure}[t]
\centering
\includegraphics[width = \linewidth]{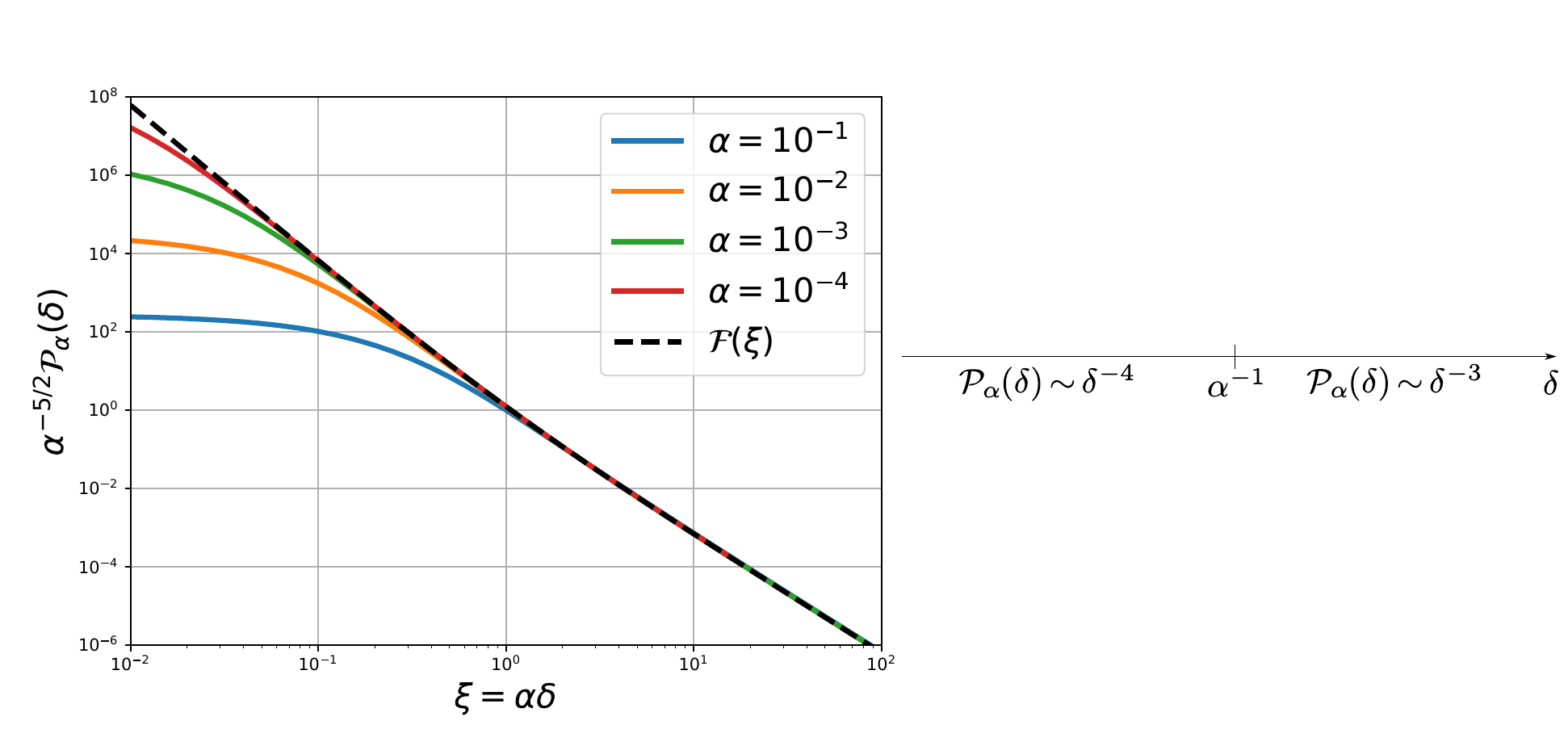}
\caption{{\bf Left}: Comparison between the rescaled scaling function $\alpha^{-5/2} {\cal P}_\alpha(\delta)$, where ${\cal P}_{\alpha}(\delta)$ is given in Eq. \eqref{P_gap} for different values of $\alpha$ and the scaling function ${\cal F}(\xi=\alpha\delta)$ given in Eq. \eqref{F}. The tail of the PDF interpolates from a $\xi^{-4}$ behaviour for $\xi\ll 1$ to a $\xi^{-3}$ behaviour for $\xi\gg 1$, matching perfectly the scaling function ${\cal F}(\xi)$. {\bf Right}: the two tail behaviours of ${\cal P}_{\alpha}(\delta)$: if $1 \ll \delta \ll \alpha^{-1}$ then ${\cal P}_\alpha(\delta) \sim \delta^{-4}$ while if $\delta \gg \alpha^{-1}$ then $P_\alpha(\delta) \propto \delta^{-3}$.}\label{Fig_F_log}
\end{figure} 
The limit $\alpha \to 0$ deserves yet another remark. Indeed, in the limit $\delta \to \infty$, the scaling function ${\cal P}_{\alpha}(\delta)$, in the bulk, has an inverse cubic tail [see the second line of Eq. (\ref{tail})], i.e.  ${\cal P}_{\alpha}(\delta) \propto \delta^{-3}$. On the other hand, at the edge (corresponding to the limit $\alpha \to 0$), the PDF of the gap decays as $P(\delta) \propto \delta^{-4}$ [see Eq. (\ref{asympt_P_edge})]. This indicates that the two limits $\alpha\to 0$ and $\delta\to \infty$ do not commute. In fact, from the full expression in (\ref{eq_p_text}) it is rather straightforward to show that there exists a scaling regime corresponding to
$\alpha\to 0$ and $\delta\to\infty$ but keeping $\xi=\alpha\delta=O(1)$ fixed, which smoothly interpolates between these two different tail behaviours.  %allowing to smoothly match these two tails in Eqs. \eqref{a_0_d_inf} and \eqref{d_inf_a_0},
In this scaling regime, we find that ${\cal P}_\alpha(\delta)$ takes the scaling form
\be\label{F}
{\cal P}_\alpha(\delta) \approx \alpha^{5/2}{\cal F}(\alpha \delta)\;\;{\rm with}\;\;{\cal F}(\xi)=\frac{2}{\pi}\frac{1}{\xi^3}+\frac{3}{\sqrt{8\pi}}\frac{1}{\xi^4}\;.
\ee
A plot of this scaling function ${\cal F}(\xi)$ together with a comparison of $\alpha^{-5/2}{\cal P}_\alpha(\delta)$ [evaluated from the exact formula in Eq. (\ref{eq_p_text})] is provided in the left panel of Fig. \ref{Fig_F_log}. This scaling form (\ref{F}) indicates that, for large $\delta \gg 1$ and small $\alpha \ll 1$, the function ${\cal P}_\alpha(\delta)$ exhibits two different tail behaviours: if $1 \ll \delta \ll \alpha^{-1}$ then ${\cal P}_\alpha(\delta) \sim \delta^{-4}$ while if $\delta \gg \alpha^{-1}$ then $P_\alpha(\delta) \propto \delta^{-3}$. This is summarised in the right panel of Fig. \ref{Fig_F_log}.

\subsection{Large deviation regime}

Since the PDF governing the typical fluctuations of the gaps has a power law tail $P_\alpha(\delta) \propto \delta^{-3}$, higher order moments of the gaps are dominated by the large deviation regime of $p_{k,n}(\Delta)$ for $\Delta = O(1)$. To study this regime, we compute the behaviour of the coefficients $A_1(z=e^{-q},s=e^{-p};\Delta)$ for $p\sim q\ll 1$ but with $\Delta=O(1)$ fixed. This yields
\begin{align}
A_1(z=e^{-q},s=e^{-p};\Delta)&\approx -\frac{1}{p}+a_2\left(p+q,p;\frac{\Delta}{\sigma}\right)+a_3\left(p+q,p;\frac{\Delta}{\sigma}\right)\\
B_1(z=e^{-q},s=e^{-p};\Delta)&\approx -\frac{1}{p+q}+a_2\left(p,p+q;\frac{\Delta}{\sigma}\right)+a_3\left(p,p+q;\frac{\Delta}{\sigma}\right)\;,
\end{align}
where the functions $a_2(r=p+q,p;\tilde \Delta)$ and $a_3(r=p+q,p;\tilde \Delta)$ read
\begin{align}
&a_2(r,p;\tilde \Delta)=\frac{\coth(\sqrt{2}\tilde \Delta)}{\sqrt{p}}+\frac{1}{\sqrt{r}\sinh(\sqrt{2}\tilde \Delta)},\\
&a_3(r,p;\tilde \Delta)=-\sqrt{\frac{r}{p}}\frac{1}{\sinh^2(\sqrt{2}\tilde \Delta)}-\coth^2(\sqrt{2}\tilde \Delta)-\left(1+\sqrt{\frac{p}{r}}\right)\frac{\cosh(\sqrt{2}\tilde \Delta)}{\sinh^2(\sqrt{2}\tilde \Delta)}\;.
\end{align}

In the limit of small $p\sim q\ll 1$, the leading contribution in the general formula \eqref{P_A_B} is again given by the terms involving the second derivatives, yielding
\begin{align}
&\tilde \pi(r,p;\Delta)\approx\frac{\sigma}{\sqrt{2}}\partial_{\Delta}^2\left(\frac{a_2(r,p;\frac{\sqrt{2}\Delta}{\sigma})+a_3(r,p;\frac{\sqrt{2}\Delta}{\sigma})}{\sqrt{p}}+\frac{a_2(p,r;\frac{\sqrt{2}\Delta}{\sigma})+a_3(p,r;\frac{\sqrt{2}\Delta}{\sigma})}{\sqrt{r}}\right).
\end{align}
In this expression, we need to keep only the terms that depend both on $r$ and $p$ as they are the only terms giving physical contribution when inverting the Laplace transforms. These terms read

\begin{align}
&\tilde \pi(r=p+q,p;\Delta)\approx\frac{\sqrt{2}}{\sigma}\left[\frac{1}{\sqrt{r p}}\frac{3+\cosh\left(\frac{2\sqrt{2}\Delta}{\sigma}\right)}{\sinh^3\left(\frac{\sqrt{2}\Delta}{\sigma}\right)}-2\left(\frac{\sqrt{r}}{p}+\frac{\sqrt{p}}{r}\right)\frac{1+\cosh\left(\frac{2\sqrt{2}\Delta}{\sigma}\right)}{\sinh^4\left(\frac{\sqrt{2}\Delta}{\sigma}\right)}\right]+\cdots\;.
\end{align}
For this large deviation form, the Laplace transforms are simple to invert, using that
\be
u^{-a}=\int_0^{\infty}e^{-ux}\frac{x^{a-1}}{\Gamma(a)}dx\;,\;\;a>0\;.
\ee
For large $n$ and large $k$ we obtain finally for $\Delta=O(1)$
\be\label{LD_gen}
p_{k,n}(\Delta)\approx\frac{1}{n\sigma}\frac{\Psi\left(\frac{\Delta}{\sigma}\right)}{\sqrt{\alpha(1-\alpha)}}+\frac{1}{\sigma n^{3/2}}\left(\frac{1}{\alpha^{3/2}}+\frac{1}{(1-\alpha)^{3/2}}\right)\varphi_0\left(\frac{\Delta}{\sigma}\right)\;, 
\ee
where $\Psi(\tilde \Delta)$ and $\varphi_0(\tilde\Delta)$ are given by
\be\label{LD_gap}
\Psi(\tilde \Delta)=\frac{\sqrt{2}}{\pi}\frac{3+\cosh(2\sqrt{2}\tilde \Delta)}{\sinh^3( \sqrt{2}\tilde \Delta)}\;, \; \varphi_0(\tilde\Delta)=\sqrt{\frac{2}{\pi}}\frac{2+\cosh(2\sqrt{2}\tilde \Delta)}{\sinh^4(\sqrt{2}\tilde \Delta)} \;.
\ee
The leading behaviour of this large deviation form (\ref{LD_gen}) will be different in the large $n$ limit, depending on whether $k=O(1)$ remains fixed or $k=O(n)$. Indeed, taking first the limit $n\to \infty$ with $k$ fixed (i.e. $\alpha \to 0$, corresponding to the edge), the dominant contribution is the term $\propto \alpha^{-3/2}$ in Eq. (\ref{LD_gap}), leading to
\be\label{Delta_1_k_1}
p_{k,n}(\Delta)\approx\frac{1}{\sigma k^{3/2}}\varphi_0\left(\frac{\Delta}{\sigma}\right)\;,\;\;\Delta=O(1)\;\; {\rm and}\;\;k=O(1)\;,
\ee
recovering the result of Ref. \cite{SM12}. 

On the other hand, in the bulk, with $n, k\to \infty$ with $\alpha=k/n$ fixed, the leading contribution is given by the term of order $O(1/n)$ in (\ref{LD_gen}). Therefore, in the bulk, the large deviation form of the PDF of the gap reads
\be\label{Delta_1_k_n}
p_{k,n}(\Delta)\approx\frac{1}{n\sigma}\frac{\Psi\left(\frac{\Delta}{\sigma}\right)}{\sqrt{\alpha(1-\alpha)}}\;,\;\;\Delta=O(1)\;\; {\rm and}\;\;k=O(n)\;.
\ee
Note that in the limit $\tilde \Delta \to 0$, the large deviation function $\Psi(\tilde \Delta)$ behaves as $\Psi(\tilde \Delta) \approx (2/\pi) \tilde{\Delta}^{-3}$, which matches smoothly with the tail behaviour of the typical regime [see the second line of Eq. (\ref{tail})].  

Finally, we end up this section on the large deviations by noting that all the terms in Eq. (\ref{LD_gen}) become of the same order in the intermediate regime where $k = O({\sqrt{n}})$. Indeed, setting  $\lambda=k/\sqrt{n}=O(1)$ one obtains, from (\ref{LD_gen}), that $p_{k,n}(\Delta)$ takes the scaling form 
\begin{align}\label{G_matching}
p_{k,n}(\Delta)\approx \frac{1}{\sigma k^{3/2}}{\cal G}\left(\frac{k}{\sqrt{n}},\frac{\Delta}{\sigma}\right)\;,\;\;{\cal G}(\lambda,\tilde \Delta)=\lambda \Psi(\tilde \Delta)+\varphi_0(\tilde \Delta)\;,
\end{align}
which smoothly interpolates between (\ref{Delta_1_k_1}) in the limit $\lambda \to 0$ and (\ref{Delta_1_k_n}) in the limit $\lambda \to \infty$.

Let us now investigate the consequences of this behaviour (\ref{LD_gen}) on the moments of the gaps. 

\subsection{Computation of the moments}

Since the scaling function ${\cal P}_{\alpha}(\delta)$ that describes the typical gap fluctuations behaves as ${\cal P}_\alpha(\delta) \sim \delta^{-3}$ for large $\delta$, the first moment of the gap is indeed completely dominated by the typical region [see Eq. (\ref{app_m1})]. This is however not the case for higher order moments. Indeed, the two non-trivial contributions to the large deviation form in Eq. \eqref{LD_gen} will both contribute to the moments of order $p\geq 2$. In the case $p>3$, we obtain
\be
\frac{\moy{d_{k,n}^p}}{\sigma^p}\approx \frac{M_p}{\sqrt{k(n-k)}}+\left(\frac{1}{k^{3/2}}+\frac{1}{(n-k)^{3/2}}\right)m_p\;,\;\;p>3\;,
\ee
where the values of $M_p$ and $m_p$ can be computed explicitly (see Appendix \ref{app:moments}),
\begin{align}
M_p&=\int_0^{\infty}dx \, x^p \, \Psi(x)=\frac{2^{\frac{4-p}{2}}}{\pi}p!(1-2^{1-p})\zeta(p-1)\;,\label{M_p}\\
m_p&=\int_0^{\infty}dx \, x^p \, \varphi_0(x)=\frac{2^{\frac{4-3p}{2}}}{\sqrt{\pi}}p!\zeta(p-2)\;,\label{m_p}
\end{align}
where $\zeta(s)=\sum_{k=1}^{\infty}k^{-s}$ is the Riemann Zeta function. The cases $p=2$ and $p=3$ are particular since there are logarithmic corrections. For $p=2$ we obtain
\be\label{m_p_2}
\frac{\moy{d_{k,n}^2}}{\sigma^2}\approx \frac{\ln n}{\pi\sqrt{k(n-k)}}+\frac{1}{2}\left(\frac{1}{k^{3/2}}+\frac{1}{(n-k)^{3/2}}\right)\;,
\ee
while, for $p=3$ we have
\be\label{m_p_3}
\frac{\moy{d_{k,n}^2}}{\sigma^3}\approx \frac{3\pi}{2\sqrt{2k(n-k)}}+\frac{3}{4\sqrt{2\pi}}\left(\frac{\ln k}{k^{3/2}}+\frac{\ln (n-k)}{(n-k)^{3/2}}\right)\;.
\ee
Finally, these behaviour can be summarised as follows
\be\label{moments}
\frac{\moy{d_{k,n}^p}}{\sigma^p}\approx \begin{cases}
\displaystyle\frac{1}{\sqrt{2\pi k}}+\frac{1}{\sqrt{2\pi(n-k)}}&\;,\;\;p=1\\
&\\
\displaystyle\frac{\ln n}{\pi\sqrt{k(n-k)}}+\frac{1}{2}\left(\frac{1}{k^{3/2}}+\frac{1}{(n-k)^{3/2}}\right)&\;,\;\;p=2\\
&\\
\displaystyle\frac{3\pi}{2\sqrt{2k(n-k)}}+\frac{3}{4\sqrt{2 \pi}}\left(\frac{\ln k}{k^{3/2}}+\frac{\ln (n-k)}{(n-k)^{3/2}}\right)&\;,\;\;p=3\\
&\\
\displaystyle\frac{M_p}{\sqrt{k(n-k)}}+\left(\frac{1}{k^{3/2}}+\frac{1}{(n-k)^{3/2}}\right)m_p&\;,\;\;p>3\;,
\end{cases}
\;n\gg 1\;,k\gg 1\;,
\ee
 In the regime $\alpha=k/n=O(1)$, the first term in the last three lines of Eq. \eqref{moments} gives the leading contribution to the moments. On the other hand, in the regime $k=O(1)$, it is the term in $k^{-3/2}$ that is dominant. Finally, in the intermediate regime mentioned above $k=O(\sqrt{n})$, both of these terms are of the same order for $p>3$ while the term with the logarithmic correction is dominant for $p=2,3$.

\section{Conclusion}\label{sec:conclusion}

In this article, we have computed exactly the PDF $p_{k,n}(\Delta)$ of the gap between two successive maxima $M_{k,n}$ and $M_{k+1,n}$ for a random walk with double exponential (Laplace) jump distribution. The main focus of the present paper has been the limiting distribution of the gap 
$d_{k,n}$ in the scaling limit where both $n$ and $k$ are large, keeping the ratio $\alpha = k/n$ fixed. This allowed us to study the gaps in the {\it bulk}, i.e. far from the global maximum $x_{\max}$ of the random walk after $n$ steps (see Fig. \ref{Fig_pp}). Our main result is an explicit expression for the distribution 
${\cal P}_{\alpha}(\delta)$ [see Eq. (\ref{P_gap})] which governs the typical fluctuations of $d_{k,n}$ in this scaling limit, namely for $d_{k,n} = O(n^{-1/2})$. We conjecture that this distribution ${\cal P}_{\alpha}(\delta)$ is universal for all random walks with a jump distribution $f(\eta)$ which is continuous, symmetric and possesses a finite second moment. What happens for heavy tailed jump distributions, i.e. the case of L\'evy flights, remains a challenging open question, in particular because we do not know how to solve the backward integral equations \eqref{inteq_1} and \eqref{inteq_2} in this case. We hope that the results obtained here will motivate further works to develop alternative methods to study the gap statistics of L\'evy flights.   

We found rather useful to think about the different positions of the random walker after $n$ steps as a point process on a line, as illustrated in Fig. \ref{Fig_pp}. By analogy with random matrices  we naturally identify edge regions, close to the extremal positions of the random walk, as well as the bulk region, far from the maximum and the minimum. In particular, we have shown that the gaps behave quite differently in these two regions. Pursuing this analogy with random matrices, one may wonder whether one can define a ``density'' associated to this point process that would capture the existence of these edge regions, and would be the equivalent of the Wigner semi-circle in random matrix theory. This is left for future investigations \cite{tbp}.

%Another interesting extension of this work would be to obtain the distribution of the gap in the case of L\'evy flights where the variance of the jump distribution is infinite. Finally, this work could be extended to study the PDF of the gap in the case where there are $M$ independent walkers.

\appendix

\section{Explicit expression for ${\cal P}_{\alpha}(\delta)$}\label{app:explicit}

The integrals in the expression \eqref{P_gap} for the PDF ${\cal P}_\alpha(\delta)$ can be expressed in terms of the following integrals
\begin{align}\label{I2_3}
&I_2(\delta,b) = \int_0^\infty y^2 e^{-\delta y - by^2} = \frac{\sqrt{\pi }}{8 b^{5/2}} e^{\frac{\delta^2}{4 b}} \left(2 b+\delta^2\right) \erfc\left(\frac{\delta}{2 \sqrt{b}}\right)-\frac{\delta}{4 b^{2}}\\
&I_3(\delta,b) = \int_0^\infty y^3 e^{-\delta y - by^2} =\frac{4 b+\delta^2}{8 b^3}-\frac{\sqrt{\pi } \delta e^{\frac{\delta^2}{4 b}} \left(6 b+\delta^2\right) \erfc\left(\frac{\delta}{2 \sqrt{b}}\right)}{16 b^{7/2}}\end{align}
as well as (see formula 4 p. 178 of Ref. \cite{Prudnikov})
\begin{align}\label{Prudnikov}
J(\delta,b,c) = \int_0^\infty e^{-\delta y} e^{-b y^2}\, {\rm erf}(c y) =& \frac{3\,c}{4 b^{5/2}} \Psi_1\left(\frac{5}{2}, \frac{1}{2}; \frac{3}{2}, \frac{1}{2}; -\frac{c^2}{b},\frac{\delta^2}{4b} \right) \\
&-\frac{2 c\, \delta }{\sqrt{\pi}\,b^3} \Psi_1\left(3, \frac{1}{2}; \frac{3}{2}, \frac{3}{2}; -\frac{c^2}{b},\frac{\delta^2}{4b} \right)
\end{align}
where $\Psi_1$ is a confluent hypergeometric series of two variables (sometimes called Humbert series \cite{wikipedia_humbert}) defined as
\begin{align}\label{humbert}
\Psi_1(a,b;c_1,c_2;x,y) = \sum_{m=0}^\infty \sum_{n=0}^\infty \frac{(a)_{m+n}\,(b)_m}{(c_1)_m(c_2)_n} \frac{x^m}{m!} \frac{y^n}{n!} 
\end{align}
where $(a)_m = \Gamma(a+m)/\Gamma(a)$ is the Pochhammer symbol. In terms of $I_2(\delta,b)$ and $I_3(\delta,b)$ in (\ref{I2_3}) and $J(\delta,b,c)$ in (\ref{Prudnikov}), the gap distribution ${\cal P}_\alpha(\delta)$ in \eqref{P_gap} reads
\begin{subequations}\label{explicit_Pa_g}
\begin{align}
&{\cal P}_\alpha(\delta) = \frac{1}{\pi \sqrt{\alpha(1-\alpha)}} \, I_2\left(\delta,\frac{1}{8 \alpha(1-\alpha)}\right) \\
&+ \frac{1}{4\sqrt{2 \pi}} \left( \frac{1}{(1-\alpha)^{\frac{3}{2}}}\, I_3\left(\delta,\frac{1}{8(1-\alpha)}\right)   +  \frac{1}{\alpha^{\frac{3}{2}}}\, I_3\left(\delta,\frac{1}{8\alpha}\right) \right) \\
& - \frac{1}{4\sqrt{2 \pi}}\left( \frac{1}{(1-\alpha)^{\frac{3}{2}}}\, J\left(\delta,\frac{1}{8(1-\alpha)},\frac{1}{2\sqrt{2\alpha}}\right)   +  \frac{1}{\alpha^{\frac{3}{2}}}\, J\left(\delta,\frac{1}{8\alpha},\frac{1}{2\sqrt{2 (1-\alpha)}}\right) \right) \;.
\end{align}
\end{subequations}
Note that this expression (\ref{explicit_Pa_g}) is explicitly symmetric under the change $\alpha \to 1-\alpha$, as it should, i.e. ${\cal P}_\alpha(\delta) = {\cal P}_{1-\alpha}(\delta)$.

In the special case $\alpha=1/2$ (which corresponds to the vicinity of the median), the integrals in Eq. \eqref{explicit_Pa_g} can be performed in terms
of elementary functions (using in particular formula 2 p. 175 of \cite{Prudnikov}). This yields
\begin{align}\label{P1/2}
{\cal P}_{\alpha = 1/2}(\delta) = & 2  e^{\delta^2} \left(2 \delta^2+3\right) \delta \left(\erfc\left(\frac{\delta}{\sqrt{2}}\right)^2-2 \erfc(\delta)\right) \nonumber \\
&-2 \sqrt{\frac{2}{\pi} } e^{\frac{\delta^2}{2}} \left(3
   \delta^2+2\right) \erfc\left(\frac{\delta}{\sqrt{2}}\right)+\frac{8}{\sqrt{\pi }} \left(\delta^2+1\right)+\frac{4}{\pi} \delta\;,
\end{align}
which is clearly different from the scaling function found at the edge \cite{SM12} (corresponding to the limit $\alpha \to 0$), given in Eq. (\ref{typ_k_order_1}). 
In particular, its asymptotic behaviours are given by
\begin{align}\label{asympt_a1/2}
{\cal P}_{\alpha=1/2}(\delta) \approx
\begin{cases}
&\dfrac{4(2-\sqrt{2})}{\sqrt{\pi}} \;, \; \delta \to 0 \\
&\\
&\dfrac{4}{\pi} \delta^{-3} \;, \;\;\;\;\;\;\;\; \delta \to \infty \;,
\end{cases}
\end{align}
which is fully consistent with the behaviours given in Eq. (\ref{tail}) specified for $\alpha = 1/2$.

\section{Computation of $\moy{d_{k,n}}$ from ${\cal P}_\alpha(\delta)$}\label{m_1_P_a}

From Eq. \eqref{eq_p_text}, the mean value of the gap is obtained as
\be
\frac{\sqrt{n}\moy{d_{k,n}}}{\sigma}\approx \int_0^{\infty}\delta \,{\cal P}_\alpha(\delta) \, d\delta\;.
\ee
Using that $\int_0^{\infty}y^2\delta e^{-\delta y}d\delta=1$, we obtain the expression
\begin{align}
\frac{\sqrt{n}\moy{d_{k,n}}}{\sigma}\approx &\int_{0}^{\infty}dy\frac{e^{-\frac{y^{2}}{8\alpha(1-\alpha)}}}{\pi\sqrt{\alpha(1-\alpha)}}+\int_{0}^{\infty}dy \frac{ye^{-\frac{y^{2}}{8(1-\alpha)}}}{4\sqrt{2\pi}(1-\alpha)^{\frac{3}{2}}}\erfc\left(\frac{y}{2\sqrt{2\alpha}}\right)\nn\\
&+\int_{0}^{\infty}dy\frac{y e^{-\frac{y^{2}}{8\alpha}}}{4\sqrt{2\pi}\alpha^{\frac{3}{2}}}\erfc\left(\frac{y}{2\sqrt{2(1-\alpha)}}\right)\;.\label{moy_d}
\end{align}
The second integral is identical to the third under $\alpha\to 1-\alpha$. It can be computed using an integration by part,
\bea
&&\int_{0}^{\infty} \frac{dy \,y \, e^{-\frac{y^{2}}{8(1-\alpha)}}}{4\sqrt{2\pi}(1-\alpha)^{\frac{3}{2}}}\erfc\left(\frac{y}{2\sqrt{2\alpha}}\right) \\
&&=\left[-\frac{e^{-\frac{y^{2}}{8(1-\alpha)}}}{\sqrt{2\pi(1-\alpha)}}\erfc\left(\frac{y}{2\sqrt{2\alpha}}\right)\right]_0^{\infty}-\int_0^{\infty}\frac{e^{-\frac{y^2}{8\alpha(1-\alpha)}}dy}{2\pi\sqrt{\alpha(1-\alpha)}}\;,
\eea
where we used that $\erfc(x)=\frac{2}{\sqrt{\pi}}\int_x^{\infty}dz e^{-z^2}$. 
Note that the last term of this equation allows to simplify half of the first term of Eq. \eqref{moy_d}. As there are two such terms (coming from the integrals with $\alpha$ and $1-\alpha$), the final result reads
\be\label{app_m1}
\frac{\sqrt{n}\moy{d_{k,n}}}{\sigma}\approx\frac{1}{\sqrt{2\pi}}\left(\frac{1}{\sqrt{\alpha}}+\frac{1}{\sqrt{1-\alpha}}\right)=\mu(\alpha)\;,
\ee
where we used that $\erfc(0)=1$ and $\erfc(x \to \infty) = 0$.
\section{Computations of $M_p$ and $m_p$}\label{app:moments}
To obtain the value of $M_p$, we compute the moment of order $p$ of the large deviation scaling function of the PDF $\Psi(\tilde \Delta)$. Using that $\Psi(x)=(\sqrt{2}/\pi)\partial_{x}^2 (\sinh^{-1}(\sqrt{2}x))$, $M_p$ reads after integration by part
\be
M_p=\int_0^{\infty}dx \, x^p \,\Psi(x)=\frac{\sqrt{2}}{\pi}p(p-1)\int_0^{\infty}dx \frac{x^{p-2}}{\sinh(\sqrt{2}x)}\;.
\ee
Changing variable from $x \to z=\sqrt{2}x$, we obtain
\be
M_p=\frac{p(p-1)2^{\frac{2-p}{2}}}{\pi}\int_0^{\infty}dz \frac{z^{p-2}}{\sinh(z)}\;.
\ee
Finally, using the integral representation of the Riemann Zeta function \cite{zeta_1},
\be
\zeta(s)=\frac{1}{2(1-2^{-s})\Gamma(s+1)}\int_0^{\infty}dx\frac{x^{s-1}}{\sinh(s)}\;,
\ee
we obtain the final result in Eq. \eqref{M_p}.

To obtain the value of $m_p$, we proceed similarly, computing the moment of order $p$ of the large deviation scaling function of the PDF $\varphi_0(\tilde \Delta)$. Using that $\varphi_0(x)=(8\pi)^{-1/2}\partial_x^2 (\sinh^{-2}(x))$, $m_p$ reads after integration by part
\be
m_p=\int_0^{\infty}dx \, x^p \,\varphi_0(x)=\frac{p(p-1)}{\sqrt{8\pi}}\int_0^{\infty}dx \frac{x^{p-2}}{\sinh^2(\sqrt{2}x)}.
\ee
Changing variable $x\to z=\sqrt{2}x$, we obtain
\be
m_p=\frac{p(p-1)}{2^{\frac{p+2}{2}}\sqrt{\pi}}\int_0^{\infty}dz\frac{z^{p-2}}{\sinh^2(z)}\;.
\ee
Finally, using the integral representation \cite{zeta_2} of the Riemann Zeta function,
\be
\zeta(s)=\frac{2^{s-1}}{\Gamma(s+1)}\int_0^{\infty}dx\frac{x^s}{\sinh^2(s)}\;,
\ee
we obtain the final expression in Eq. \eqref{m_p}.
{}

\end{document}